\RequirePackage{fix-cm}
\documentclass[twocolumn,epjc3]{svjour3}  
\smartqed  

\RequirePackage[T1]{fontenc}
\RequirePackage{graphicx}
\RequirePackage{mathptmx}      
\RequirePackage[numbers,sort&compress]{natbib}
\RequirePackage[colorlinks,citecolor=blue,urlcolor=blue,linkcolor=blue]{hyperref}

\usepackage{multirow}
\usepackage{xspace}
\usepackage{amsmath}
\usepackage{amssymb}
\usepackage[utf8]{inputenc}
\usepackage[switch]{lineno}
\usepackage{enumerate}
\RequirePackage{graphicx}
\usepackage{color}
\usepackage{xspace}
\journalname{Eur. Phys. J. C}

\newcommand{\VTPCOne}{{VTPC-1}\xspace}
\newcommand{\VTPCTwo}{{VTPC-2}\xspace}
\newcommand{\GAPTPC}{{GAP-TPC}\xspace}

\newcommand{\MTPCL}{{MTPC-L}\xspace}
\newcommand{\MTPCR}{{MTPC-R}\xspace}
\newcommand{\BPDOne}{{BPD-1}\xspace}
\newcommand{\BPDTwo}{{BPD-2}\xspace}
\newcommand{\BPDThree}{{BPD-3}\xspace}
\newcommand{\BPDAll}{{BPD-1/2/3}\xspace}

\newcommand{\dedx}{\ensuremath{{\rm d}E\!/\!{\rm d}x}\xspace}
\newcommand{\pt}{\ensuremath{p_\mathrm{T}}\xspace}
\newcommand{\mt}{\ensuremath{m_\mathrm{T}}\xspace}
\newcommand{\sincol}[1]{\multicolumn{1}{r@{\hspace{2pt}}}{#1}}
\DeclareMathOperator\atanh{atanh} 

\newcommand{\GeV}{\ensuremath{\mbox{Ge\kern-0.1em V}}\xspace}
\newcommand{\GeVc}{\ensuremath{\mbox{Ge\kern-0.1em V}\!/\!c}\xspace}
\newcommand{\GeVcc}{\ensuremath{\mbox{Ge\kern-0.1em V}\!/\!c^2}\xspace}
\newcommand{\AGeV}{\ensuremath{A\,\mbox{Ge\kern-0.1em V}}\xspace}
\newcommand{\AGeVc}{\ensuremath{A\,\mbox{Ge\kern-0.1em V}\!/\!c}\xspace}
\newcommand{\UrqmdLong}{{\scshape U}r{\scshape qmd3.4}\xspace}
\newcommand{\Urqmd}{{\scshape U}r{\scshape qmd}\xspace}
\newcommand{\GeantThree}{{\scshape Geant3}\xspace}
\newcommand{\Epos}{{\scshape Epos}\xspace}
\newcommand{\EposLong}{{\scshape Epos1.99}\xspace}

\newcommand{\HsdLong}{{\scshape Hsd2.0}\xspace}
\newcommand{\Fritiof}{{\scshape Fritiof}\xspace}
\newcommand{\FritiofLong}{{\scshape Fritiof7.02}\xspace}
\newcommand{\NASixtyOne}{\mbox{NA61\slash SHINE}\xspace}

\title{Production of $\Lambda$-hyperons in inelastic p+p interactions at 158~\GeVc
}

\institute{{National Nuclear Research Center, Baku, Azerbaijan}\label{inst0}
\and{Faculty of Physics, University of Sofia, Sofia, Bulgaria}\label{inst1}
\and{Ru{\dj}er Bo\v{s}kovi\'c Institute, Zagreb, Croatia}\label{inst2}
\and{LPNHE, University of Paris VI and VII, Paris, France}\label{inst3}
\and{Karlsruhe Institute of Technology, Karlsruhe, Germany}\label{inst4}
\and{Fachhochschule Frankfurt, Frankfurt, Germany}\label{inst5}
\and{University of Frankfurt, Frankfurt, Germany}\label{inst6}
\and{University of Athens, Athens, Greece}\label{inst7}
\and{Wigner Research Centre for Physics of the Hungarian Academy of Sciences, Budapest, Hungary}\label{inst8}
\and{Institute for Particle and Nuclear Studies, Tsukuba, Japan}\label{inst9}
\and{University of Bergen, Bergen, Norway}\label{inst10}
\and{Jan Kochanowski University in Kielce, Poland}\label{inst11}
\and{National Centre for Nuclear Research, Warsaw, Poland}\label{inst12}
\and{Jagiellonian University, Cracow, Poland}\label{inst13}
\and{University of Silesia, Katowice, Poland}\label{inst14}
\and{University of Warsaw, Warsaw, Poland}\label{inst15}
\and{University of Wroc{\l}aw,  Wroc{\l}aw, Poland}\label{inst16}
\and{Warsaw University of Technology, Warsaw, Poland}\label{inst17}
\and{Institute for Nuclear Research, Moscow, Russia}\label{inst18}
\and{Joint Institute for Nuclear Research, Dubna, Russia}\label{inst19}
\and{National Research Nuclear University ``MEPhI'' (Moscow Engineering Physics Institute), Moscow, Russia}\label{inst20}
\and{St. Petersburg State University, St. Petersburg, Russia}\label{inst21}
\and{University of Belgrade, Belgrade, Serbia}\label{inst22}
\and{ETH Z\"urich, Z\"urich, Switzerland}\label{inst23}
\and{University of Bern, Bern, Switzerland}\label{inst24}
\and{University of Geneva, Geneva, Switzerland}\label{inst25}
\and{Los Alamos National Laboratory, Los Alamos, USA}\label{inst27}
\and{University of Colorado, Boulder, USA}\label{inst28}
\and{University of Pittsburgh, Pittsburgh, USA}\label{inst29}
\and{{\it Present address:} Department of Physics, COMSATS Institute of Information Technology, Islamabad 44000 Pakistan}\label{inst29}
\\
$^\mathsection$ Corresponding author: \url{Tatjana.Susa@irb.hr} 
}
\author{
{A.~Aduszkiewicz}\thanksref{inst15}
\and{Y.~Ali}\thanksref{inst13}
\and{E.~Andronov}\thanksref{inst21}
\and{T.~Anti\'ci\'c}\thanksref{inst2}
\and{N.~Antoniou}\thanksref{inst7}
\and{B.~Baatar}\thanksref{inst19}
\and{F.~Bay}\thanksref{inst23}
\and{A.~Blondel}\thanksref{inst25}
\and{M.~Bogomilov}\thanksref{inst1}
\and{A.~Brandin}\thanksref{inst20}
% ...
\\(\NASixtyOne Collaboration) 
}

\date{Received: date / Revised version: date}

\begin{document}
\sloppy
\maketitle

\abstract{
Inclusive production of \mbox{$\Lambda$-hyperons} was measured 
with the large acceptance \NASixtyOne spectrometer at the CERN SPS
in inelastic p+p interactions at beam momentum of 158~\GeVc. 
Spectra of transverse momentum and transverse mass
as well as distributions of rapidity and x$_{_F}$ 
are presented. The mean multiplicity was estimated to be 
$0.120\;\pm0.006\;(stat.)\;\pm0.010\;(sys.)$.
The results are compared with previous measurements and  
predictions of the \Epos, \Urqmd and \Fritiof models. 
}


\section{Introduction}
\label{intro}
Hyperon production in proton-proton (p+p) interactions
has been studied in a long series of fixed target and collider experiments. 
However, the resulting experimental data suffers from low statistics,
incomplete beam momentum coverage, and large differences between the
measurements reported by different experiments.
Also popular models of proton-proton interactions mostly fail to
reproduce the measurements.
The data on $\Lambda$ production and the model predictions are reviewed at the end of this paper.

At the same time rather impressive progress was made in measurements of
hyperon production in nucleus-nucleus (A+A) collisions~\cite{Blume:2011sb}. 
This has two reasons. 
Firstly, mean multiplicities of all hadrons in central heavy ion collisions 
are typically two to
three orders of magnitude higher than the corresponding multiplicities
in inelastic p+p interactions. Secondly, the hyperon yields per nucleon 
are enhanced by substantial factors in A+A collisions with
respect to p+p interactions. This enhancement, which increases with
the strangeness
content of the hyperon in question, has raised considerable interest over
the past decades. It has in particular been brought into connection with
production of the Quark-Gluon Plasma, a 'deconfined' state of matter 
at that time hypothetical~\cite{Rafelski:1982pu,Rafelski:1986}.
Nowadays, for the energies well below the LHC energy range, nucleus-nucleus collisions are
investigated mainly to find the critical point of strongly interacting matter
as well as to study the properties of the 
onset of deconfinement~\cite{Gazdzicki:2010iv,Gazdzicki:2015ska}.
In particular, precise measurements of inclusive hadron production 
properties as a function of beam momentum (13$A$-158$A$~\GeVc)
and  size of colliding nuclei (p+p, p+Pb, Be+Be, Ar+Sc, Xe+La) 
are performed by \NASixtyOne~\cite{Facility}. 
Results on inelastic p+p interactions are an important part
of this scan.


\NASixtyOne already published results on $\pi^{\pm}$,  K$^{\pm}$, proton, $\Lambda$ and  $K^0_S$  production in p+C interactions at beam momentum of 31~\GeVc~\cite{NA61pCPions, NA61pCKaons, NA61pCLambda, NA61thinTarget}, as well as $\pi^-$ production in p+p collisions at 20-158~\GeVc~\cite{NA61ppPions}.

 This paper presents the first \NASixtyOne  results on strange particle
 production in p+p interactions. Since all $\Sigma^0$ hyperons decay
 electromagnetically via $\Sigma^0\rightarrow\Lambda\gamma$,
 which is indistinguishable from direct $\Lambda$ production,
 $\Lambda$ in the following denotes the sum of both $\Lambda$ directly
 produced in
 strong p+p interactions and $\Lambda$ from decays of $\Sigma^0$ hyperons
 produced in these interactions.

The particle rapidity is calculated in the collision centre of
mass system (cms): $y = \atanh(\beta_\mathrm{L})$,
where $\beta_\mathrm{L} = p_\mathrm{L}/E$ is the longitudinal component of the
velocity,
$p_\mathrm{L}$ and $E$ are longitudinal momentum and energy
in the cms and $x_\mathrm{F} = p_\mathrm{L}/p_{beam}$ is
Feynman's scaling variable with $p_{beam}$ the incident proton momentum in the cms.
The transverse component of the momentum is denoted as \pt and
the transverse mass \mt is defined as $\mt = \sqrt{m^2 + \pt^2}$,
where $m$ is the particle mass.
The collision energy per nucleon pair in the centre of mass
system is denoted as $\sqrt{s_\mathrm{NN}}$.

\section{The experimental setup}
\label{sec:setup}
\begin{figure*}
\begin{center}
\resizebox{0.85\textwidth}{!}{
  \includegraphics{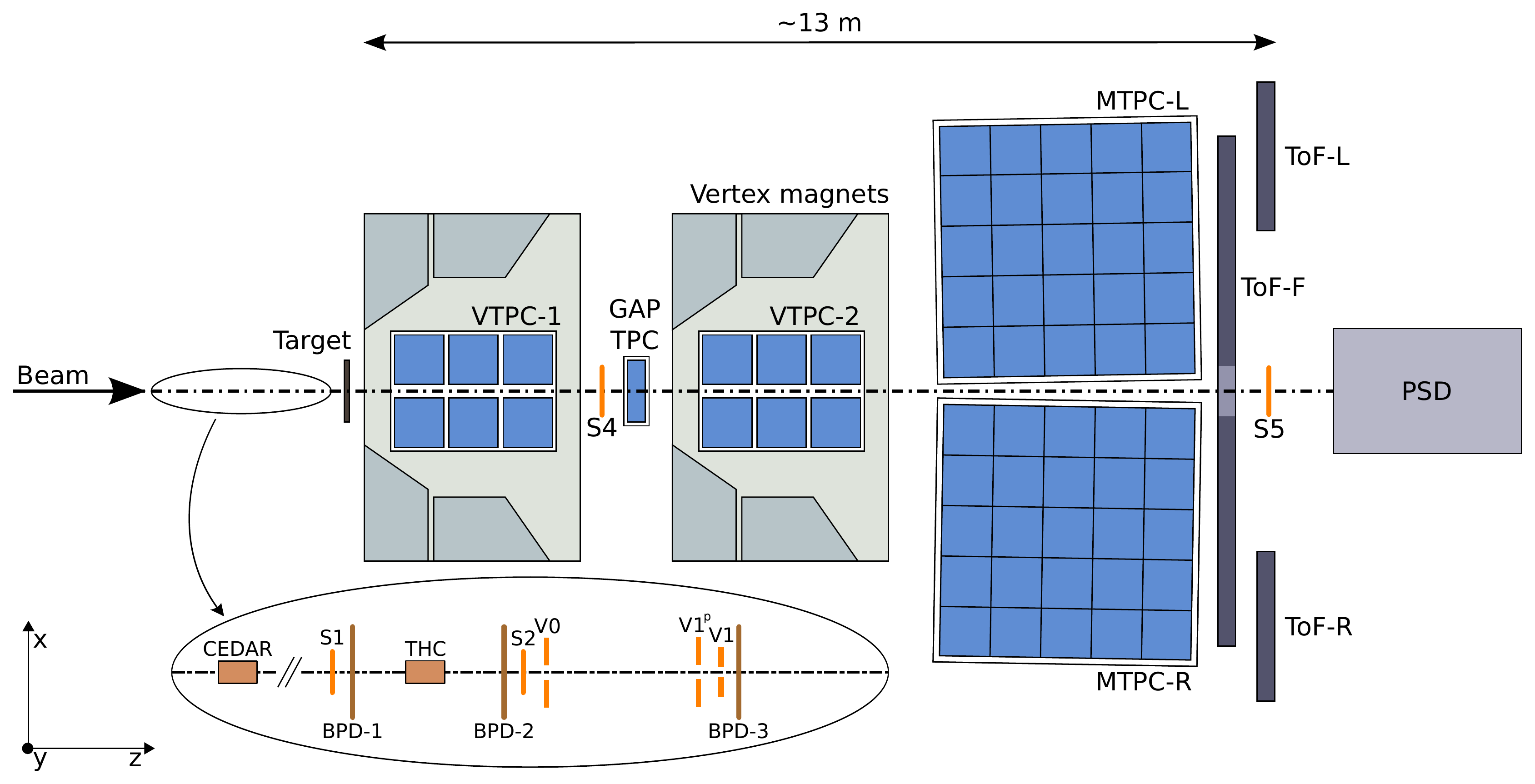}
}
\end{center}
\caption{Schematic layout of the NA61/SHINE experiment at the CERN SPS
(horizontal cut in the beam plane, not to scale).
The beam and trigger counter configuration used for data taking
on p+p interactions in 2009 is presented.
The chosen
right-handed coordinate system is shown on the plot.
The incoming beam direction is along the z axis.
The magnetic field bends charged particle trajectories in the x-z
(horizontal) plane.
The drift direction in the TPCs is along the y (vertical) axis~\cite{Facility}. 
See details in Section~\ref{sec:setup}.}
\label{fig:na61} 
\end{figure*}

The \NASixtyOne experiment~\cite{Facility} uses a large 
acceptance hadron spectrometer located
in the H2 beam-line at the CERN SPS accelerator complex. The layout of the
experiment is schematically shown in Fig.~\ref{fig:na61}. 
Hereby we describe only the components relevant for the analysis. The main detector system 
is a set of large volume Time Projection Chambers (TPCs).
Two of them (\mbox{\VTPCOne} and \mbox{\VTPCTwo}) are placed 
inside super-conducting magnets
(\mbox{VTX-1} and \mbox{VTX-2})
with a combined bending power of 9~Tm. The standard current setting for
data taking
at 158~\GeVc corresponds to full field, 1.5~T, in the first and reduced field,
1.1~T, in the second magnet.
Two large TPCs (\MTPCL and \MTPCR) are positioned downstream of the magnets, symmetrically to the undeflected beam.
A fifth small TPC (\GAPTPC) is placed between \mbox{\VTPCOne} 
and \mbox{\VTPCTwo} directly
on the beam line and
covers the gap between the sensitive volumes of the other TPCs. The \NASixtyOne
TPC system allows
a precise measurement of the particle momenta $p$ with a resolution
of $\sigma(p)/p^2\approx (0.3 - 7)\!\times\!10^{-4}\;\rm{(GeV/c)}^{-1}$ 
at the full magnetic field used for data taking at 158~\GeVc and provides
particle identification
via the measurement of the specific energy loss, \dedx, with relative
resolution of about 4.5\%.

A set of scintillation and Cherenkov counters, as well as beam position
detectors (BPDs) upstream of the main detection system provide a timing
reference, as well as identification and position measurements of the incoming beam
particles. The 158~\GeVc secondary hadron beam was produced by 
400~\GeVc primary
protons impinging on a 10~cm long beryllium target. Hadrons produced at the
target are transported downstream to the \NASixtyOne experiment by the
H2 beamline, in which collimation and momentum selection occur. Protons
from the secondary hadron beam are identified by a differential Cherenkov
Counter (CEDAR) \cite{CEDAR}.  Two scintillation counters, S1 and S2,
together with the three veto counters V0, V1 and V1$^p$ were used to select
beam particles. Thus, beam particles were required to satisfy the coincidence
$S1\cdot S2\cdot\overline{V0}\cdot\overline{V1}\cdot\overline{V1^p}\cdot
$CEDAR in order to become accepted as a valid proton. Trajectories of individual 
beam particles were measured in a telescope of beam position detectors placed along 
the beam line (\BPDAll in Fig.~\ref{fig:na61}). These are multiwire proportional chambers
with two orthogonal sense wire planes and cathode strip readout, allowing
to determine the transverse coordinates of the individual beam particle at
the target position with a resolution of about 100~$\mu$m.
For data taking on p+p interactions a liquid hydrogen target (LHT) 
of 20.29~cm length (2.8\% interaction length) and 3~cm diameter 
was placed 88.4~cm
upstream of \VTPCOne. 
Data taking with inserted and removed liquid hydrogen (LH) in the LHT was
alternated in order to calculate a data-based correction for interactions with
the material surrounding the liquid hydrogen. Interactions in the target are
selected by requiring an anti-coincidence of the selected beam protons with the signal
from a small scintillation counter of 2~cm diameter (S4) placed on the beam trajectory
between the two spectrometer magnets. Further details on the experimental
setup, beam and the data acquisition can be found in~Ref.~\cite{Facility}.

\section{Analysis technique}

In the following section the analysis technique is described, starting with
the event reconstruction followed by the event and V$^0$ selections. Next the $\Lambda$ signal extraction and 
the calculation of $\Lambda$-yields are presented. Then the correction procedure and the estimation of 
statistical and systematic uncertainties are discussed. Finally quality tests are performed on the final results. More details can be found in Ref.~\cite{Wilczek}.

\subsection{Track and main vertex reconstruction}

The main steps of the track and vertex reconstruction procedure are:
\begin{enumerate}[(i)]
  \item cluster finding in the TPC raw data, calculation of the cluster
  centre-of-gravity and total charge,
  \item reconstruction of local track segments in each TPC separately,
  \item matching of track segments into global tracks,
  \item track fitting through the magnetic field and determination of track
  parameters at the first measured TPC cluster,
  \item determination of the interaction vertex using the beam trajectory ($x$
  and $y$ coordinates) fitted in the BPDs and the trajectories of tracks
  reconstructed in the TPCs  ($z$ coordinate),
  \item matching of ToF hits with the TPC tracks.
\end{enumerate}

\subsection{Event selection}
\label{subs:eventCuts}
A total of $3.5\times 10^{6}$ events recorded with the LH inserted (denoted I)
and $0.43\times10^6$ with the LH removed
from the target (denoted R) were used for the analysis.
The two configurations were realised by filling the target vessel
with LH and emptying it.

Interaction events were selected by the following requirements:

\begin{enumerate}[(i)]

\item no off-time beam particle was detected 1~$\mu$s before and after the
trigger particle,

\item the trajectory of the beam particle was measured in at least one of \BPDOne
or \BPDTwo and in the \BPDThree detector and was well reconstructed (\BPDThree
is positioned close to and upstream of the LHT),

\item the fit of the $z$-coordinate of the primary interaction vertex converged
and the fitted $z$ position is found within $\pm 40$ cm of the centre of the LHT.
\end{enumerate}

The number of events after these selections
($N^I=1.66\times10^6$ for the LH inserted configuration of the target,
$N^R=43\times10^3$ for the LH removed) is treated as the raw number of
recorded inelastic events.

\subsection{V$^0$ reconstruction and selection}
\label{subs:candidateCuts}

$\Lambda$ hyperons are identified by reconstructing their decay topology
$\Lambda \rightarrow p + \pi^{-}$ (branching ratio 63.9\%). In the first
step pairs were formed from all measured positively and negatively charged
particles. V$^0$ candidates were required to have a distance of closest 
approach (dca, Fig.~\ref{fig:bxyDef}) between the two trajectories
of less than 1 cm anywhere between the position of the first measured points
on the tracks and the primary vertex. In the second step, the position
of the secondary vertex and the momenta of the decay tracks were fitted
by performing a 9-parameter $\chi^{2}$ fit employing the Levenberg-Marquardt
fitting procedure \mbox{\cite{Levenberg,Marquardt}}. In the fit the fitted
secondary vertex was added as the first point to the tracks at which the
momenta were recalculated. Finally, for each candidate
the invariant mass was calculated assuming proton (pion) mass for positively
(negatively) charged particles. To ensure a good momentum determination and
reduce the combinatorial background from random pairs,
a set of quality cuts was imposed:

\begin{figure}
\begin{center}
\resizebox{0.49\textwidth}{!}{
  \includegraphics{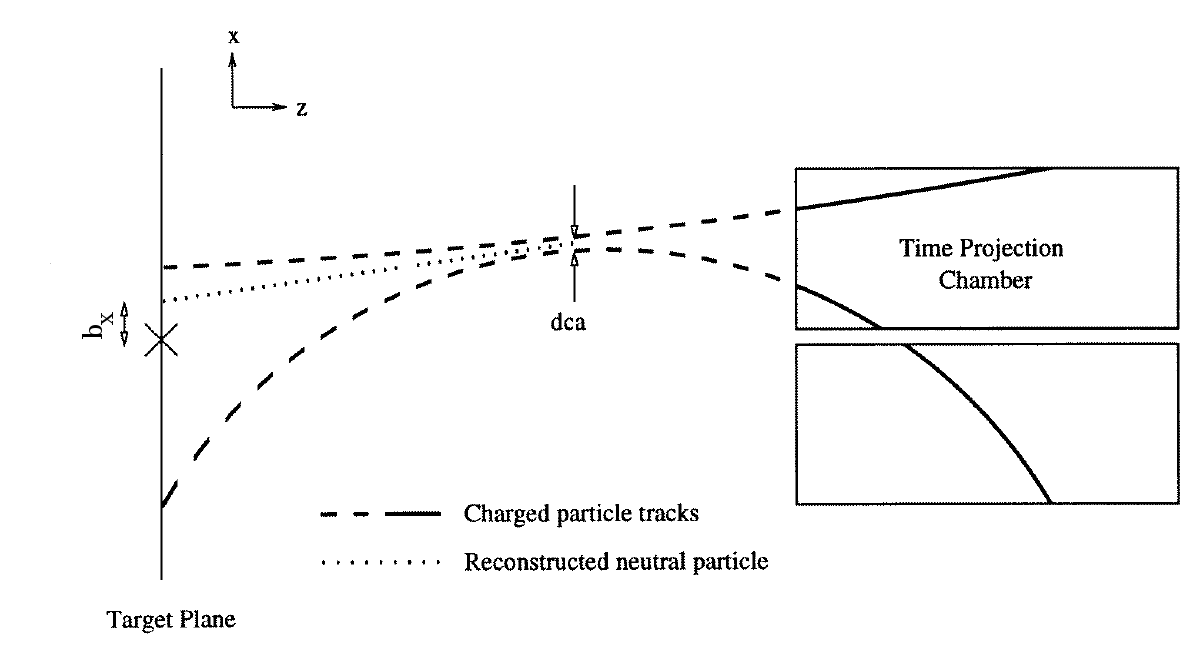}
}
\end{center}
\caption{Definition of distance of the closest approach (dca), and $b_x$. The variable $b_y$ is defined on the yz-plane in analogy with $b_x$. 
The target plane is defined as the plane parallel to the xy-plane containing the main vertex marked with a cross 
(taken from Ref. \cite{Yates}). } 
\label{fig:bxyDef}       
\end{figure}

\begin{figure}
\begin{center}
\resizebox{0.45\textwidth}{!}{
  \includegraphics{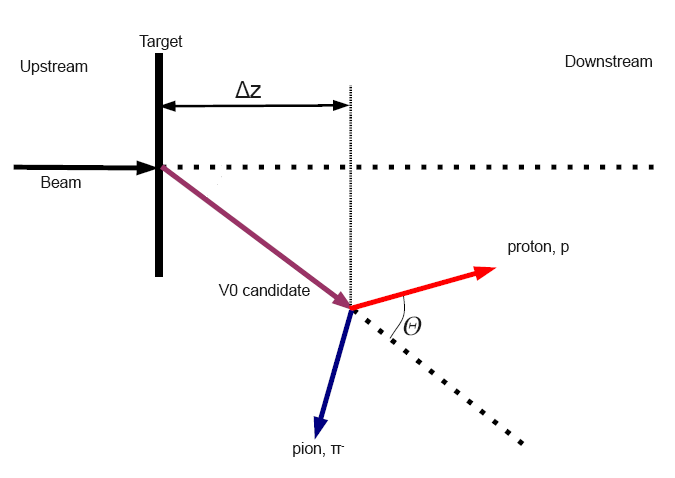}
}
\end{center}
\caption{Definition of $\Delta z$ variable used for the $V^{0}$ selection (see text).}
\label{fig:thetaDef}       
\end{figure}

\begin{figure}
\begin{center}
\resizebox{0.45\textwidth}{!}{
  \includegraphics{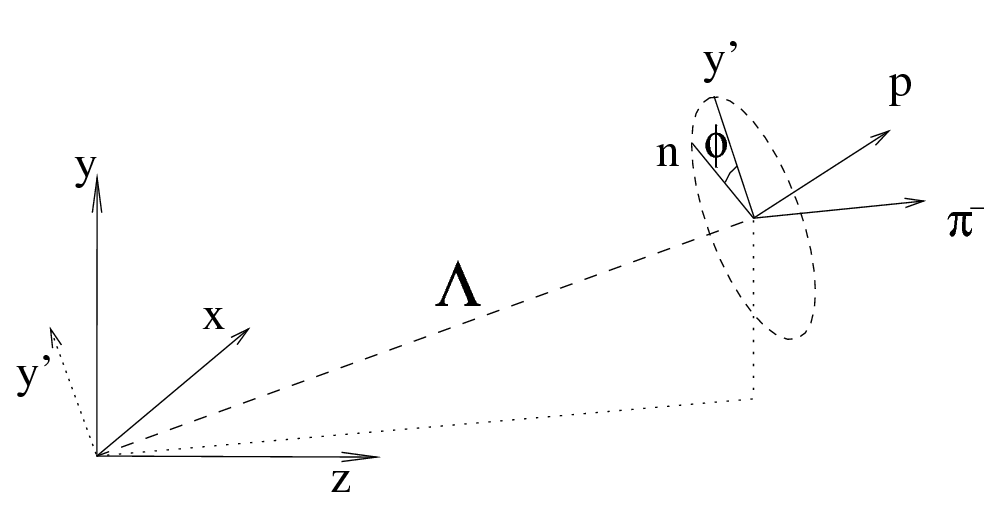}
}
\end{center}
\caption{Definition of $\phi$-variable used for the $V^{0}$ selection (see text).} 
\label{fig:phiDef}       
\end{figure}

\begin{enumerate}[(i)]
\item For each track, the minimum number of clusters in at least one of \VTPCOne and \VTPCTwo
was required to be 15.
\item Proton and pion candidates were selected by requiring their specific
energy loss measured by the TPCs to be within 3 $\sigma$ around the
nominal Bethe-Bloch value.
This cut was applied only to experimental data. 
\item For the simulated data (see below) the background was totally discarded by matching,
i.e. by using only those reconstructed tracks which were identified as
originating from the corresponding $\Lambda$ decay. The identification
was performed by matching the clusters found in the TPCs with the clusters
generated in the simulation. In case more than one reconstructed track
was matched to a $\Lambda$ decay daughter the one with the largest 
number of matched clusters was selected.
\item The combinatorial background concentrated in the vicinity of the
primary vertex is reduced by imposing a distance cut on the difference
between the $z$ coordinate of the primary and $\Lambda$ vertex ($\Delta z =
z_\Lambda - z_{primary}$, see Fig. \ref{fig:thetaDef}). To maximise the fraction of rejected background
while minimising the number of lost $\Lambda$ candidates, a rapidity 
dependent cut was applied: $\Delta z>10$ cm for $y<0.25$, $\Delta z>15$ cm
for $y\in[0.25,0.75]$, $\Delta z>40$ cm for $y\in[0.75,1.25]$, and $\Delta
z>60$ cm for higher rapidities.
\item A further significant part of the background (e.g. pairs from photon conversions) 
was rejected by imposing a cut on $\cos{\phi}$,  where $\phi$ is defined as the angle between the vectors 
$y'$, and $n$, where $y'$ is the vector perpendicular to the momentum of the V$^0$-particle 
which lies in the plane spanned by the y-axis and the V$^0$-momentum vector, 
and $n$ is a vector normal to the
decay plane (see Fig. \ref{fig:phiDef}). A rapidity dependent cut
was used: $|\cos{\phi}|<0.95$ for $y<-0.25$, $|\cos{\phi}|<0.9$ for
$y\in[-0.25,0.75]$, $|\cos{\phi}|<0.8$ for higher rapidities.
\item The trajectories of the $\Lambda$ candidates were calculated using
the decay vertex and the momentum vectors of the decay particles.
Extrapolation back to the primary vertex plane resulted in impact parameters
$b_{x}$ (in the magnetic bending plane) and $b_{y}$ (see Fig.~\ref{fig:bxyDef}). 
As the resolution of impact parameters is approximately
twice better in $y$ than in $x$ direction, an elliptic cut $\sqrt{(b_{x}/2)^2+b_{y}^{2}}< 1$ cm
was imposed in order to reduce the background from  $\Lambda$ candidates 
which do not originate from the primary vertex.
\end{enumerate}

\begin{figure}
\begin{center}
\resizebox{0.5\textwidth}{!}{
  \includegraphics{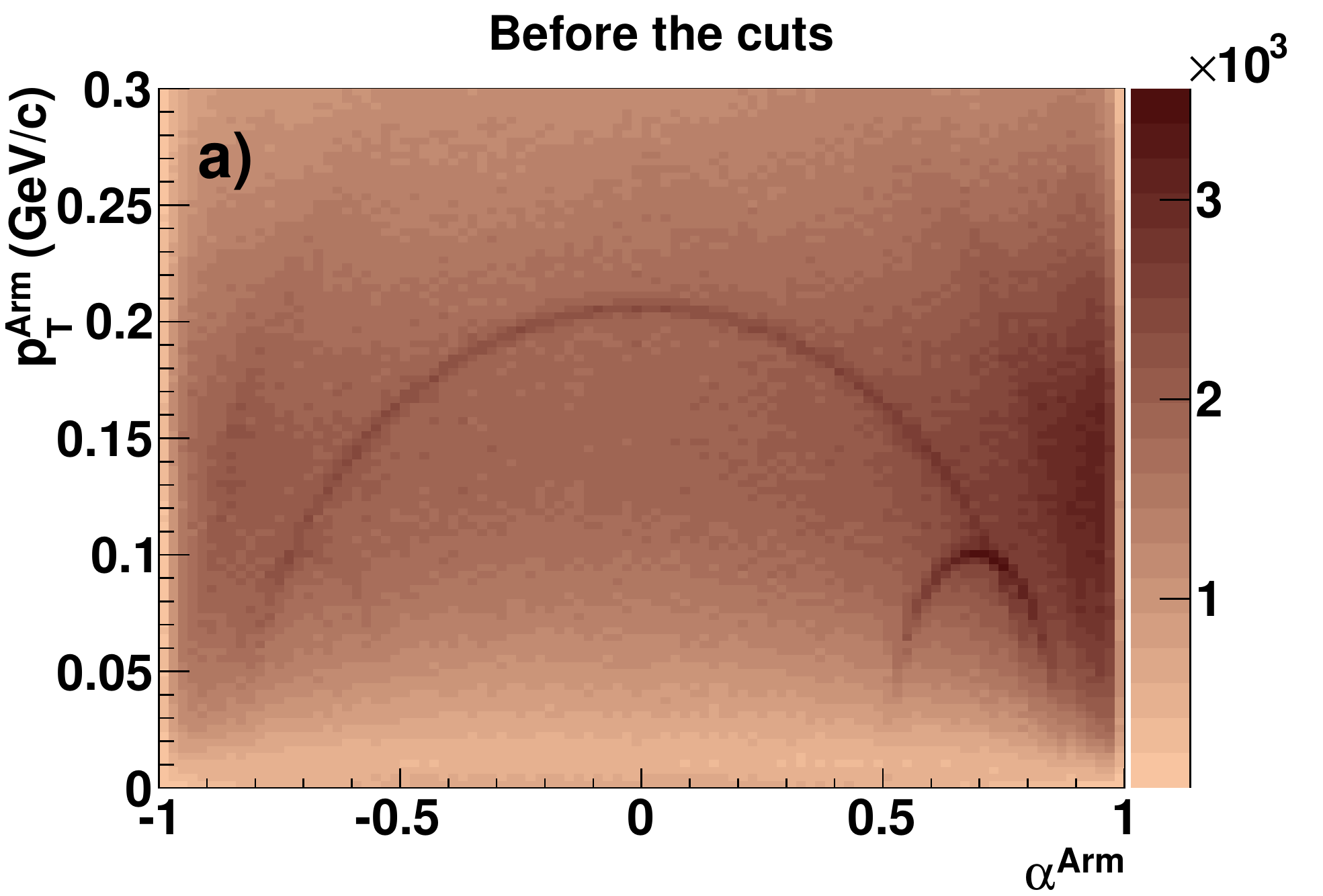}
}
\resizebox{0.5\textwidth}{!}{
  \includegraphics{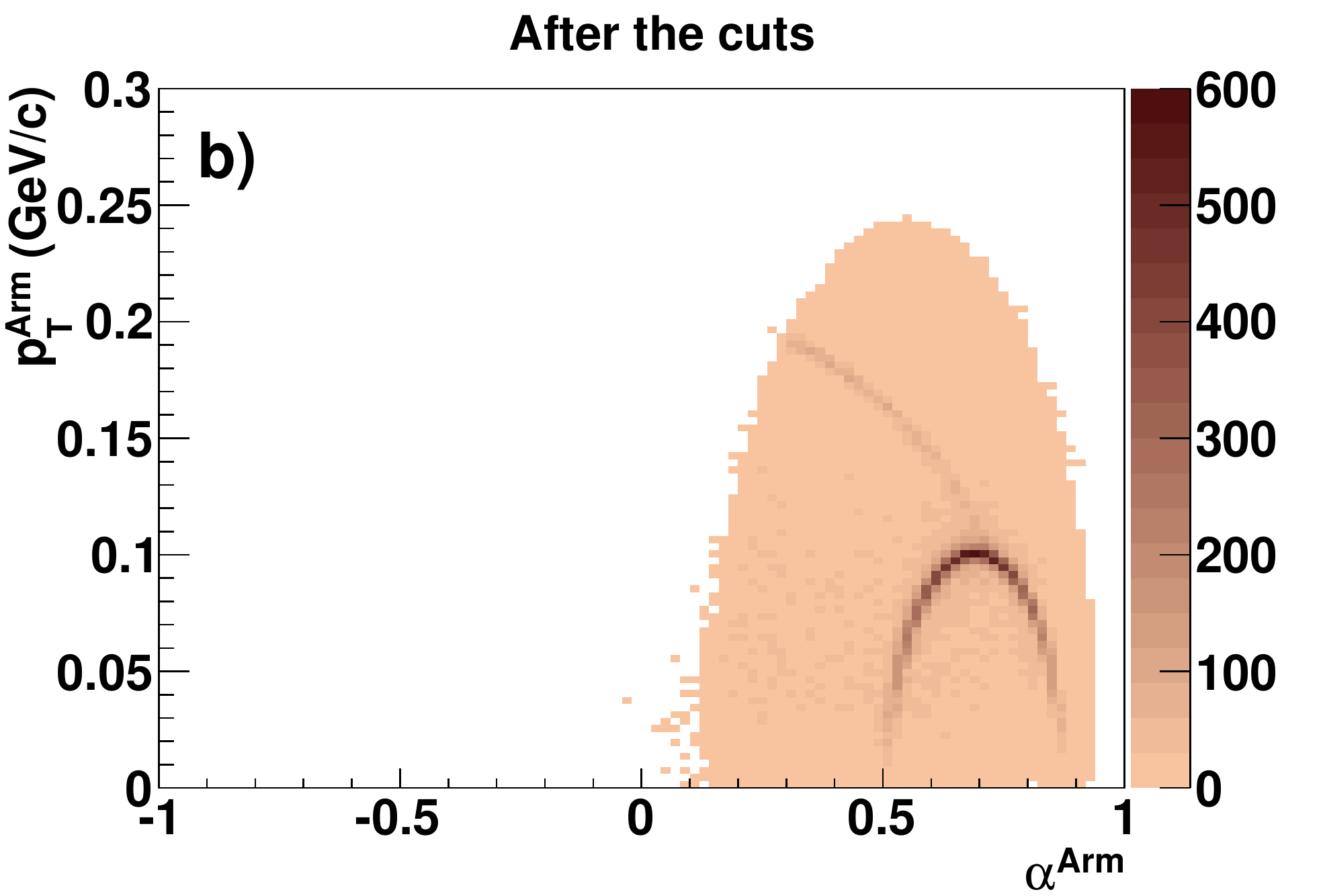}
}
\end{center}
\caption{Armenteros-Podolanski plot for reconstructed $V^0$ decays before the $\Lambda$ candidate 
selection cuts (a), and after the cuts are applied (b). The shading indicates the number of entries 
per bin. The axis variables are:  $p_T^{Arm}$,  the
transverse momentum of the decay particles with respect to the direction of motion of the $V^0$
and $\alpha^{Arm} = (p_L^+ - p_L^-)/(p_L^+ + p_L^-)$ where $p_L^+$ and $p_L^-$ are the longitudinal
momenta of the positive and negative decay particle respectively. 
}
\label{fig:armenteros} 
\end{figure}

The selection cuts lead to a high degree of purification of the $\Lambda$ signal.
This is demonstrated by the Armenteros-Podolanski plots~\cite{ArmPod} of Fig. \ref{fig:armenteros}
in which the $\Lambda$ decays populate the ellipses on the lower right.

\subsection{Signal extraction}
\label{massfit}

The raw yield of $\Lambda$ hyperons was obtained by performing a fit
of the invariant mass spectra with the sum of a background and a signal function.
The shape of the $\Lambda$ signal was described by the Lorentzian function:
\begin{equation}
L(m)=A\frac{\frac{1}{2}\Gamma}{\left(m-m_0\right)^2+\left(\frac{1}{2}\Gamma\right)^2},
\end{equation}
where $m$ is the invariant mass of the candidate ($p\pi^-)$ pair,
$A$ is a normalisation factor, $m_0$ is the mass parameter 
and $\Gamma$ is the
FWHM (full width at half maximum) of the $\Lambda$ peak. As the natural width
of $\Lambda$ decay is negligible, the observed width of the $\Lambda$ peak
is caused almost solely by the detector response. In the standard approach,
the background was represented by a Chebyshev polynomial of 2$^{nd}$ order.
The uncertainty introduced by choosing this particular functional form was
estimated by  trying other background functions (see Sec.~\ref{subs:systematics}).

\begin{figure}
\begin{center}
\resizebox{0.5\textwidth}{!}{
  \includegraphics{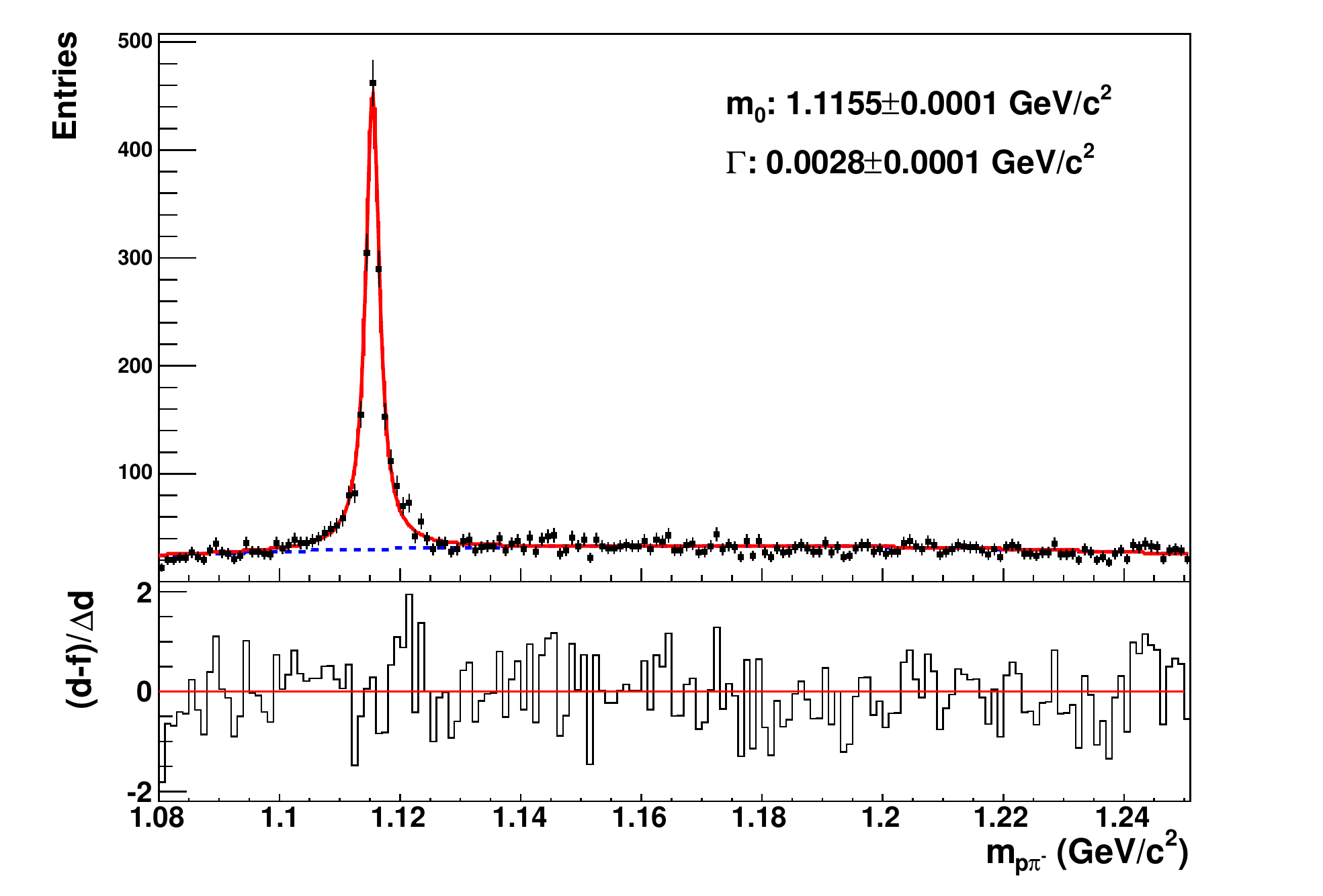}
}
\end{center}
\caption{The invariant mass distribution of $\Lambda$ candidates for
$y\in[-0.75,-0.25]$ and $p_{_T}\in[0.2,0.4]$ \GeVc is shown in the upper plot (LH inserted). 
The solid line shows a fit to signal plus background, while the dashed line
represents the background contribution. The lower part of the plot shows the difference
between the data points and the fit, normalised to the statistical error of
the data points.}
\label{fig:fit} 
\end{figure}

\begin{figure}
\begin{center}
\resizebox{0.5\textwidth}{!}{
  \includegraphics{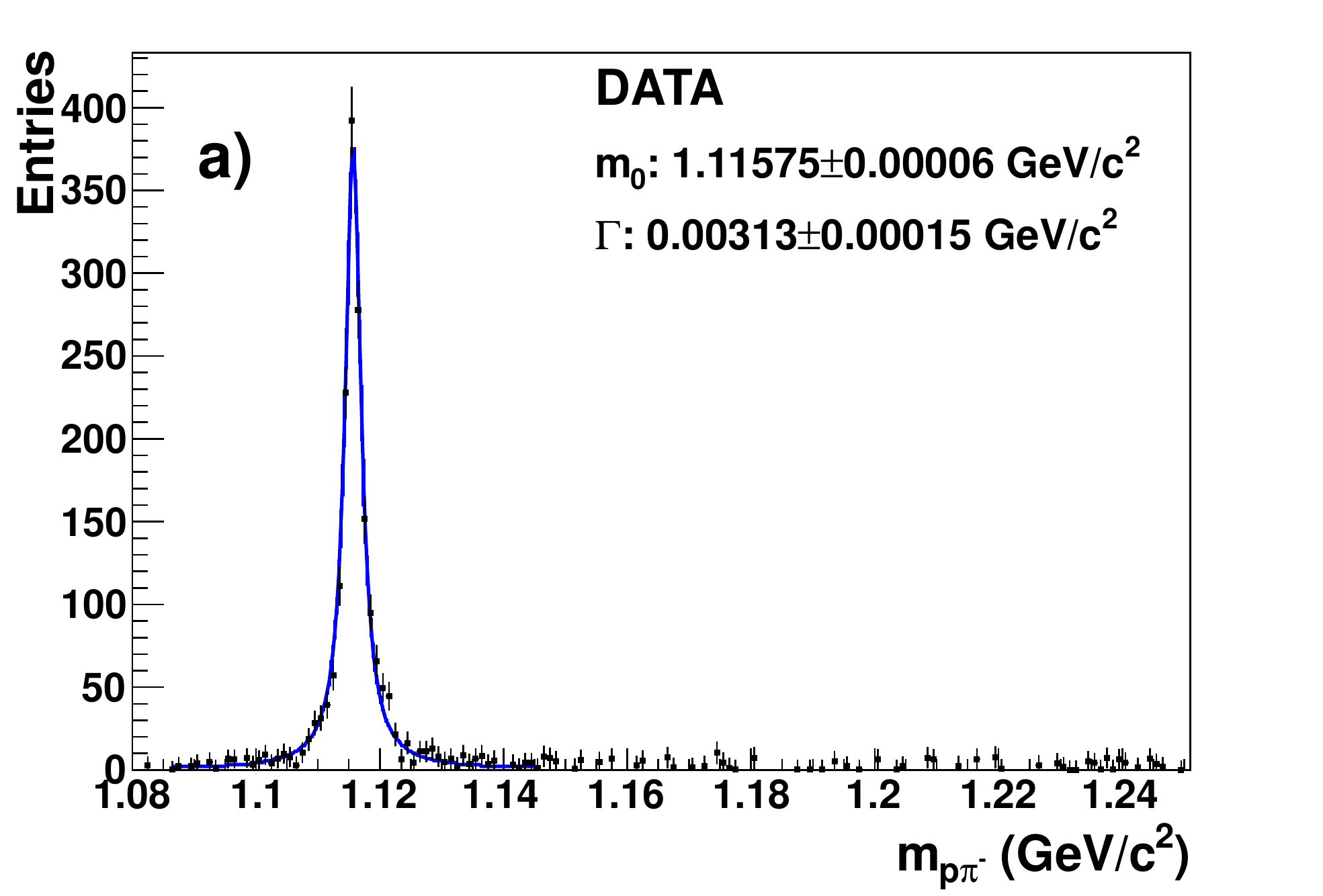}
}
\resizebox{0.5\textwidth}{!}{
  \includegraphics{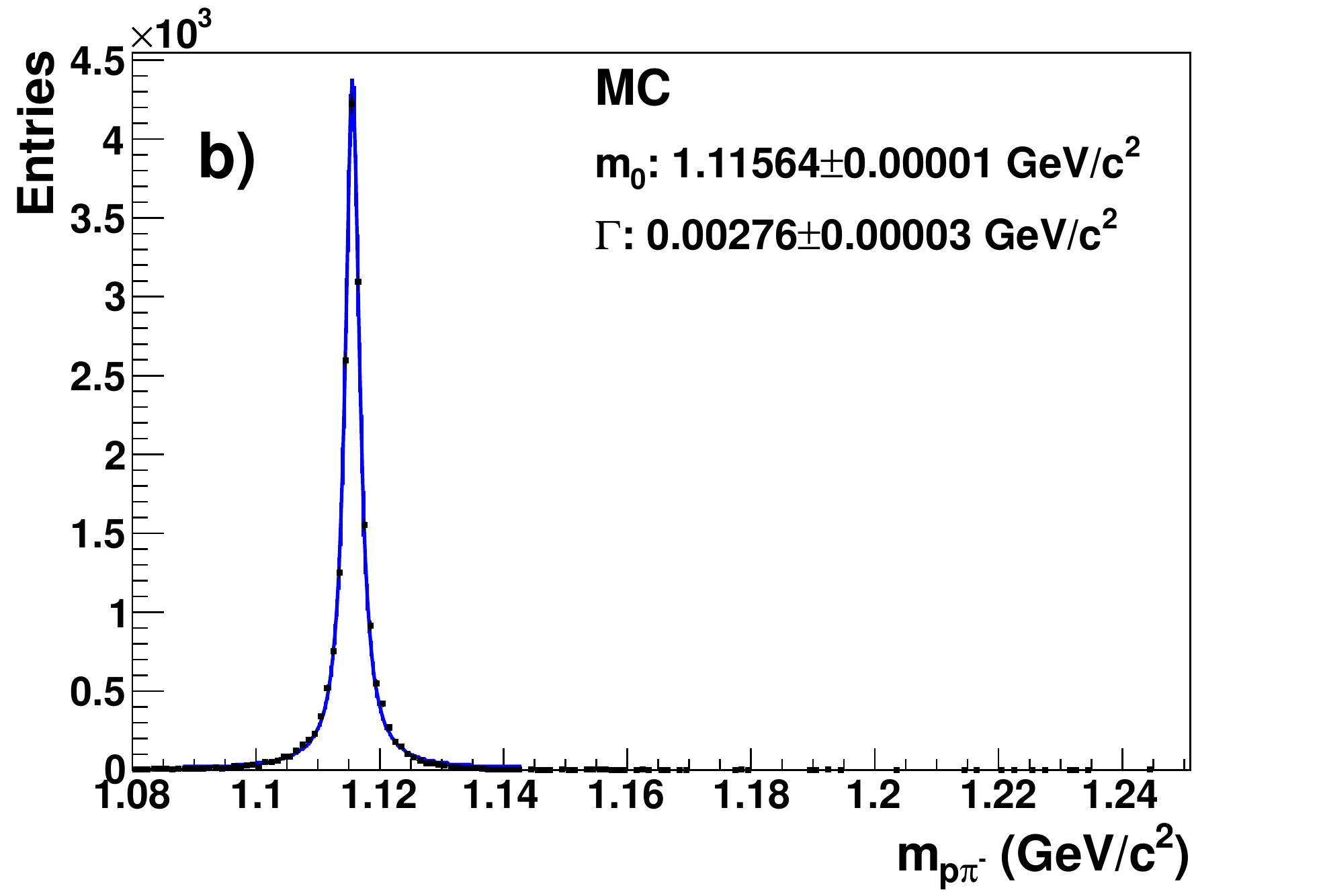}
}
\end{center}
\caption{The invariant mass distribution of $\Lambda$ candidates for
$y\in[-0.25,0.25]$ and $p_{_T}\in[0.2,0.4]$ \GeVc with the LH inserted after
subtraction of the fitted background (a), and for the simulation (b). 
}
\label{fig:fit2} 
\end{figure}

The sum of the Lorentzian and the background function was fitted in the mass range
from 1.080 (1.076 for $y=0.5$, 1.073 for $y=1.0$) to 1.250~\GeVcc.
In order to ensure the stability of the fit results, even in the case of
low statistics, a three step procedure was developed.
In the first step, a pre-fit was performed in order to estimate the initial
parameters of the background function.
For that purpose, the invariant mass region containing the $\Lambda$ peak (1.100-
1.135~\GeVcc) was excluded from the fit.
In the second step, the invariant mass spectrum was fitted to the sum of the
signal and the background. The initial values
for the parameters of the background function were taken from the first step,
while the mass parameter $m_0$ was fixed to
the PDG value $m_\Lambda=1.115683$ \GeVcc \cite{PDG} and the width was
set to 3~MeV. The obtained values were used as
the initial parameters for the third step, where no parameter was fixed. The
invariant mass distribution of the $\Lambda$ candidates
for the intervals $y\in[-0.75,-0.25]$ and $p_{_T}\in[0.2,0.4]$ \GeVc, together with
the result of the final fit is shown in Fig. \ref{fig:fit}.
For the data set with LH inserted, the fits were performed in $(k,l)$ bins,
where $k$ stands for the bin in rapidity $y$ or Feynman $x_{_F}$,
and $l$ for the bin in transverse momentum $p_{_T}$ or transverse mass
$m_{_T}-m_{_\Lambda}$. The raw number of $\Lambda$-hyperons ($n^I(k,l)$)
was then obtained by subtracting the fitted background and integrating
the remaining signal distributions in the mass window $m_0\pm3\Gamma$
(see Fig. \ref{fig:fit2}), where $m_0$ is the fitted $\Lambda$ mass.
The low statistics of the LH removed data set, forced to restrict the fits
to $y$ ($x_{_F}$) bins summed over the transverse variable, resulting in $n^R(k)$.
In order to obtain the raw number of $\Lambda$-hyperons in $(k,l)$ bins,
it was assumed that
the shape of the $p_{_T}$ distributions and the efficiencies for
a given $y$ ($x_{F}$) bin were the same for the two data sets,
and $n^R(k,l)$ was calculated as
$n^R(k,l)=n^R(k)\frac{n^I(k,l)}{\sum\limits_{l}{n^I(k,l)}}$.

\subsection{Correction factors}
\label{subs:correction}

\begin{figure}
\begin{center}
\resizebox{0.5\textwidth}{!}{
  \includegraphics{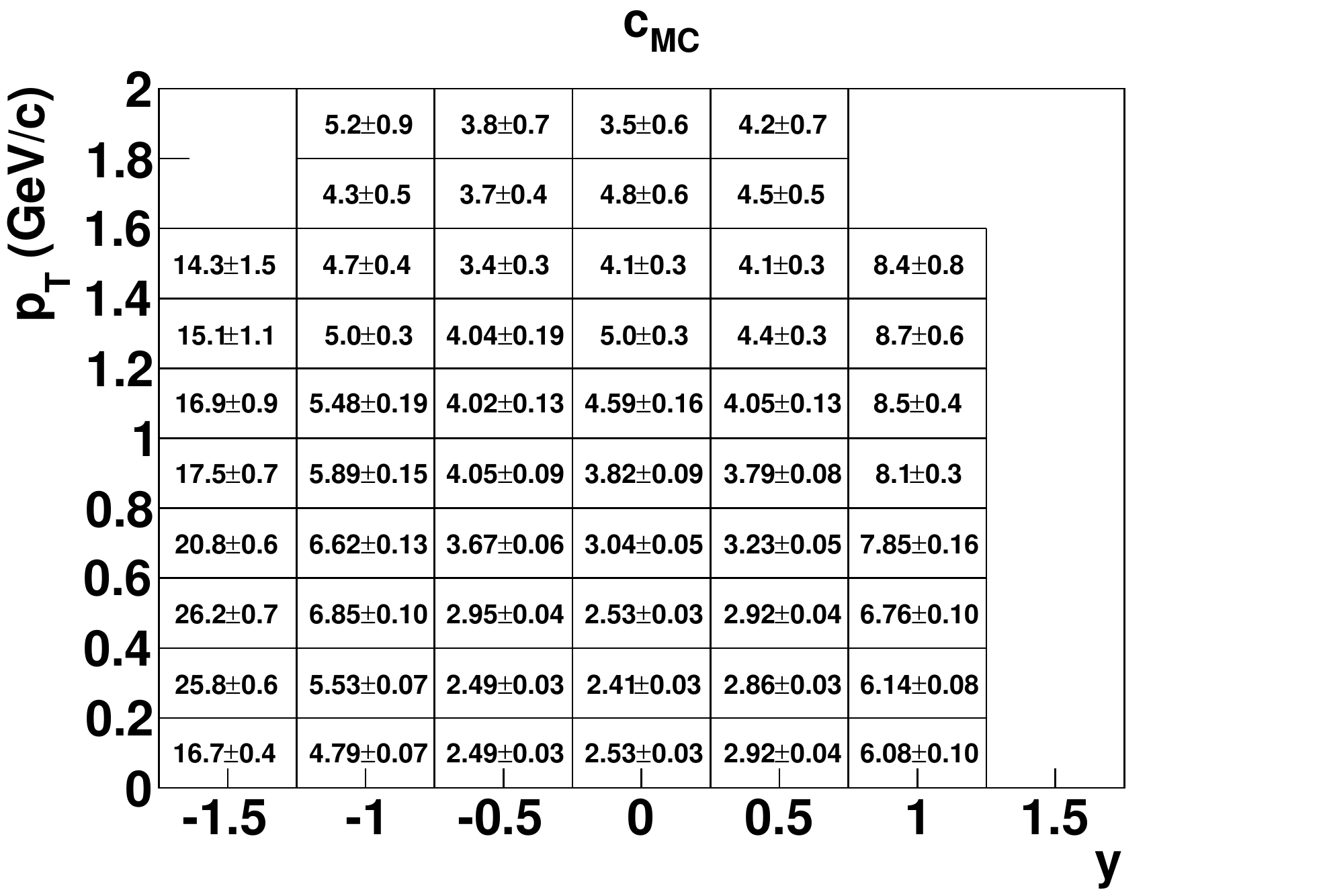}
}
\resizebox{0.5\textwidth}{!}{
  \includegraphics{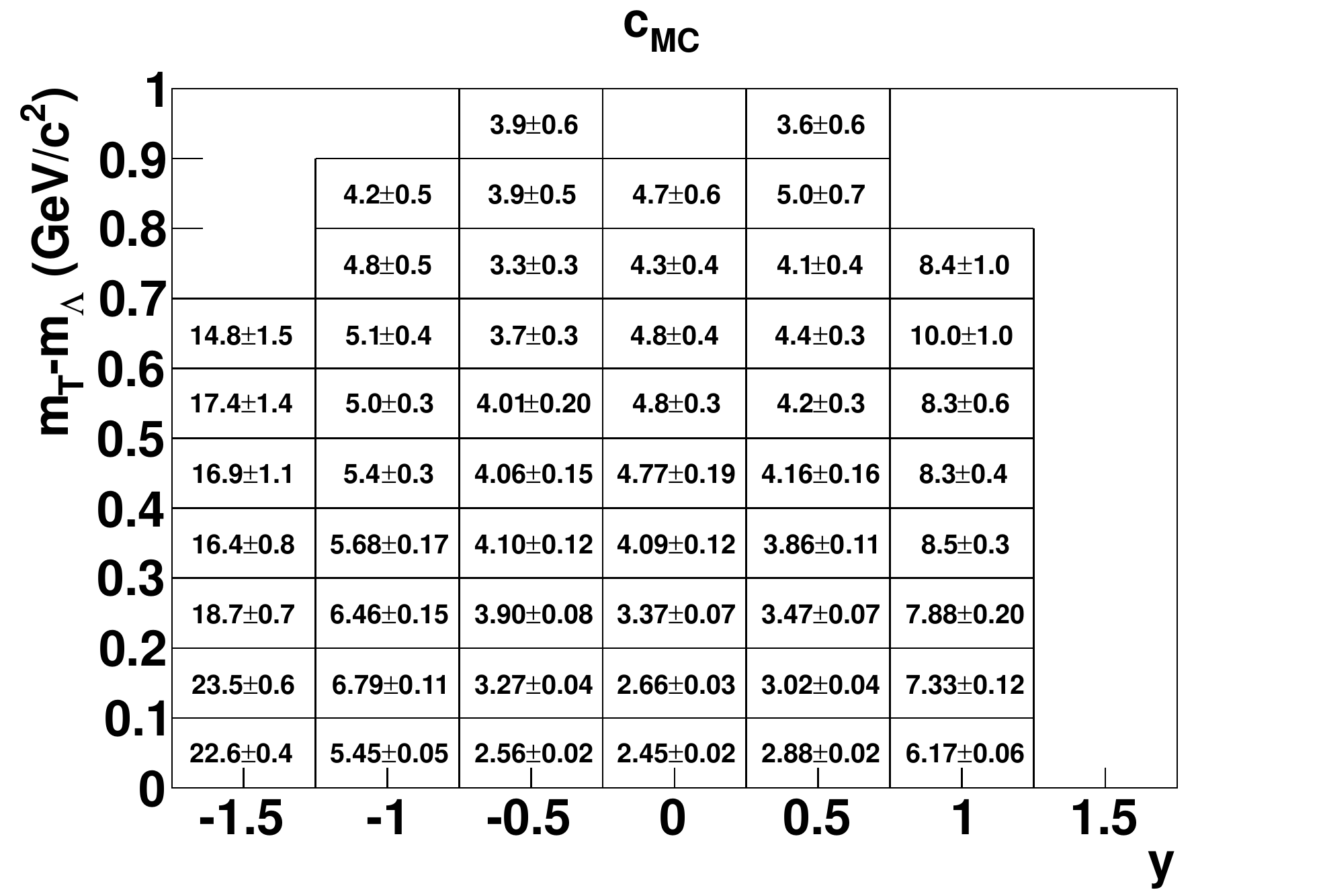}
}

\resizebox{0.5\textwidth}{!}{
  \includegraphics{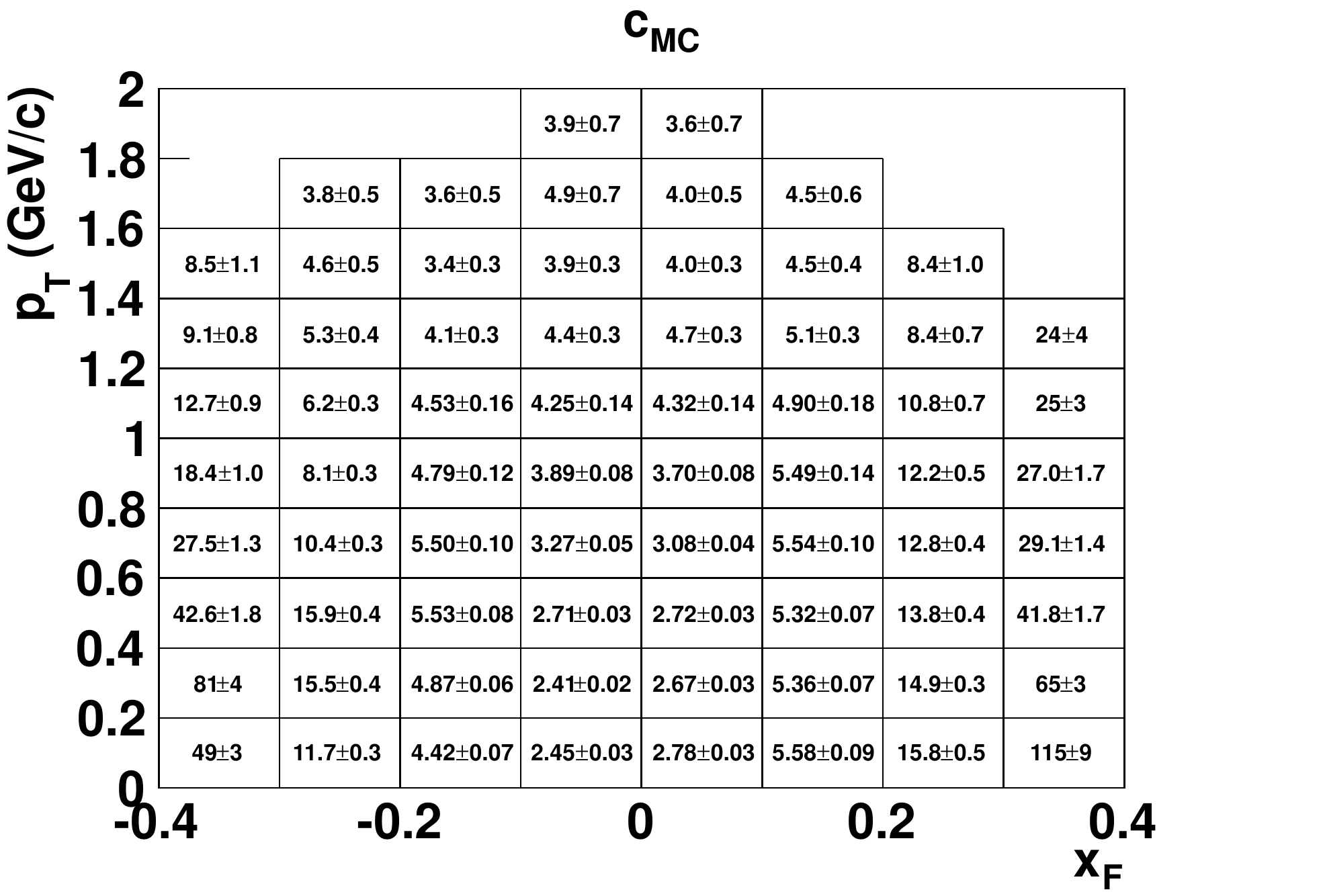}
}
\end{center}
\caption{Correction factors $c_{_{MC}}$ for binning 
in ($y,p_T$) at top, ($y,m_T-m_{\Lambda}$) at centre and ($x_F,p_T$)
at bottom. The error $\Delta c_{_{MC}}$ ranges from 0.02 to 1.48 for binning in ($y,p_T$) and ($y,m_T$),
and from 0.02 to 8.70 for ($x_F,p_T$).
 }
\label{fig:epsilon} 
\end{figure}

In order to determine the number of $\Lambda$ hyperons produced in inelastic
p+p interactions, three corrections were applied to
the extracted raw number of $\Lambda$ hyperons:
\begin{enumerate}
\item The contribution from interactions in the material outside of the liquid hydrogen
volume of the target was subtracted:
\begin{equation}
\label{eq:off-target}
\frac{n^I(k,l)-Bn^R(k,l)}{N^I-BN^R}.
\end{equation}
The normalisation factor $B$ was derived by comparing the distribution
of the fitted $z$ coordinate of the interaction vertex far away from 
the target~\cite{NA61pCLambda} for filled and empty target vessel:
\begin{equation}
\label{eq:B}
B=\frac{N^I_{far\;z}}{N^R_{far\;z}}~=3.93~,
\end{equation}
where $N^I_{far\;z}$ ($N^R_{far\;z}$) is the number of events in the region
$100 < z < 280 $ cm downstream of the target centre for the data sample with
inserted (removed) hydrogen in the target vessel. 

\item The loss of the $\Lambda$ hyperons due to the \dedx requirement,
was corrected by a constant factor
\begin{equation}
c_{\dedx} =\frac{1}{\epsilon^{2}} = 1.005~,
\end{equation}
where $\epsilon = 0.9973$ is the probability for the proton (pion) to lie 
within 3$\sigma$ around the nominal Bethe-Bloch value. \\

\item

A detailed Monte Carlo simulation was performed to correct for geometrical
acceptance, reconstruction efficiency, losses due to the
trigger bias, the branching ratio of the $\Lambda$ decay, the feed-down from hyperon decays
as well as the quality cuts applied in the analysis. The correction factors
are based on $20\times10^6$ inelastic p+p events produced
by the \EposLong event generator~\cite{EPOS}.  The particles in the generated
events were tracked through the \NASixtyOne apparatus using the \GeantThree
package~\cite{GEANT}. The TPC response was simulated by dedicated \NASixtyOne
software packages which take into account all known detector effects. The
simulated events were reconstructed with the same software as used for
real events and the same selection cuts were applied (except the identification cut).
As seen from Fig.~\ref{fig:fit2}  the shape
and position of the $\Lambda$ peak is well reproduced by the simulation while
the width is about 10\% narrower. 
More details on MC validation can be found in Ref. \cite{NA61ppPions}.

For each  $(k,l)$ bin, the correction factor $c_{_{MC}}(k,l)$
was calculated as

\begin{equation}
c_{_{MC}}(k,l)=\frac{n_{MC}^{gen}(k,l)}{N_{MC}^{gen}}\Bigg/
\frac{n_{MC}^{acc}(k,l)}{N_{MC}^{acc}}~,
\end{equation}
where
\begin{itemize}

\item $n_{MC}^{gen}(k,l)$ is the number of $\Lambda$ hyperons produced in
a given $(k,l)$ bin in the primary interactions, including $\Lambda$ hyperons
from the $\Sigma^0$ decays,
\item $n_{MC}^{acc}(k,l)$ is the number of reconstructed $\Lambda$ hyperons
in a given $(k,l)$ bin, determined by matching the reconstructed $\Lambda$
candidates to the simulated $\Lambda$ hyperons based on the cluster positions,
\item $N_{MC}^{gen}$ is the number of generated inelastic p+p interactions
($19\:961\times10^3$),
\item $N_{MC}^{acc}$ is the number of accepted p+p events
($15\:607\times10^3$),
\item $k=y$ or $x_{_F}$, and $l=p_{_T}$ or $m_{_T}-m_{_\Lambda}$.
\end{itemize}

These factors also include the correction for feed-down from weak
decays (mostly of $\Xi^{-}$ and $\Xi^{0}$, see Fig.~\ref{fig:feeddown}). The $\Xi^{-}$ yields
as function of rapidity generated by the \EposLong simulation agree
within 10\% with the measurements reported in Ref.~\cite{Tanja}. 
The values of the correction factors are presented in Fig.~\ref{fig:epsilon}.

\begin{figure}
\begin{center}
\resizebox{0.47\textwidth}{!}{
  \includegraphics{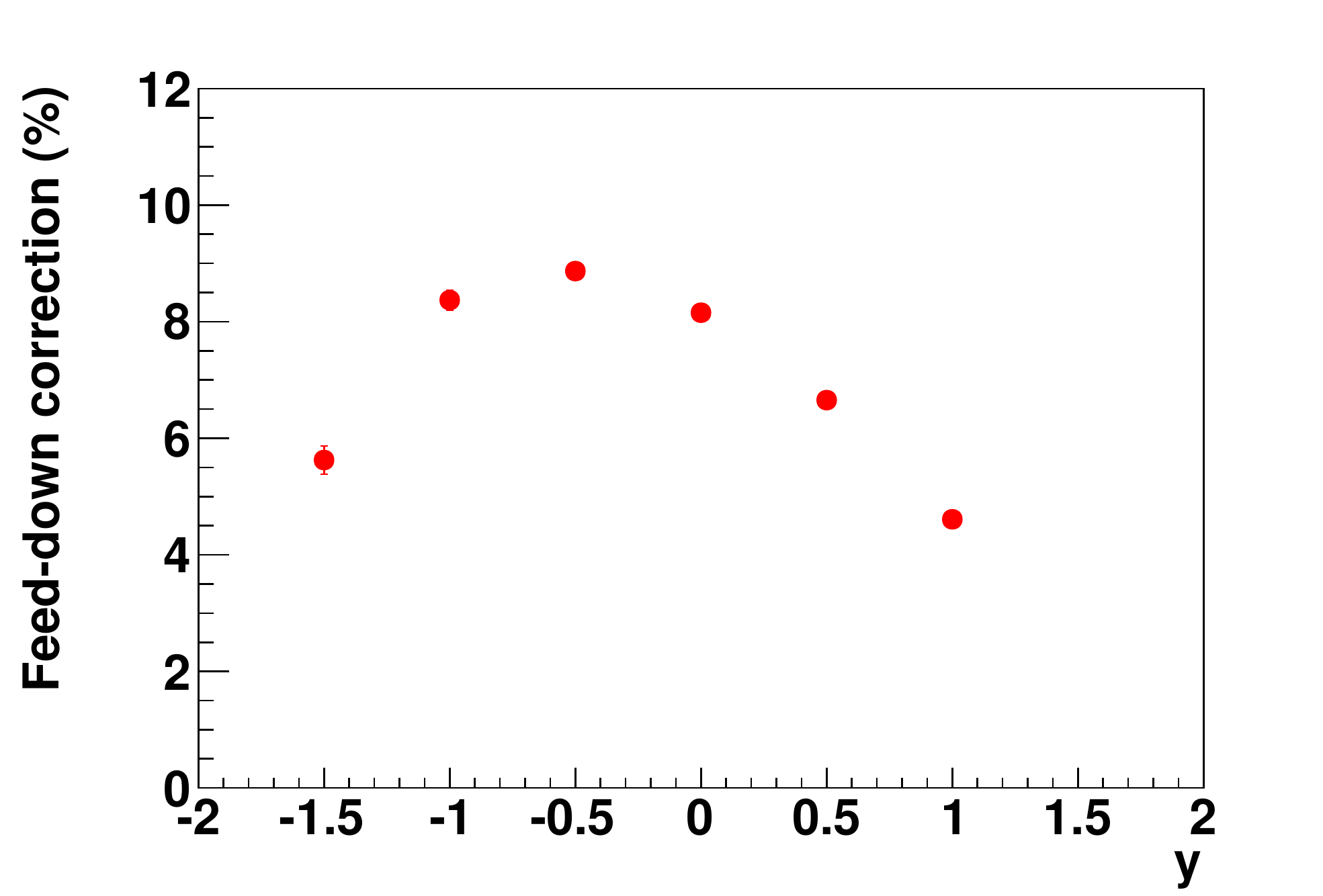}
}
\end{center}
\caption{Feed-down correction: the contribution from $\Xi^-$ and $\Xi^0$ to total $\Lambda$ decays calculated using \EposLong model.
}
\label{fig:feeddown} 
\end{figure}

Statistical errors of the~correction factors were calculated using the~following
approach:
The correction factor ($c_{_{MC}}$) consists of two parts:
\begin{equation}\begin{aligned}c_{_{MC}}\left(k,l\right)&=\frac{n_{MC}^{gen}(k,l)}
{N_{MC}^{gen}}\Bigg/\frac{n_{MC}^{acc}(k,l)}{N_{MC}^{acc}}\\
&=\frac{N^{acc}_{MC}}{N^{gen}_{MC}}\Bigg/\frac{n^{acc}_{MC}(k,l)}
{n^{gen}_{MC}(k,l)}=\frac{\alpha}{\beta(k,l)}~,
\end{aligned}\end{equation}
where $\alpha$ describes the~loss of inelastic events due to the~event
selection, and $\beta$ takes into account the~loss of $\Lambda$
hyperons due to the~$V^0$-cuts, efficiency, and the~other aforementioned
effects.

The error of $\alpha$ was calculated assuming a binomial distribution, while the~part
$\beta$ involving the~fitting procedure takes into account the error of the~fit:
\begin{equation}\Delta
\alpha=\sqrt{\frac{\alpha(1-\alpha)}{N_{MC}^{gen}}},\end{equation}
 \begin{equation}\Delta \beta(k,l)=\sqrt{\left(\frac{\Delta
 n^{acc}_{MC}(k,l)}{n^{gen}_{MC}(k,l)}\right)^2+
 \left(\frac{n^{acc}_{MC}(k,l)\Delta
 n^{gen}_{MC}(k,l)}{(n^{gen}_{MC}(k,l))^2}\right)^2}~,
\end{equation}
 where $\Delta n^{acc}_{MC}(k,l)$ is the~uncertainty of the~fit,
 and $\Delta n^{gen}_{MC}(k,l)=\sqrt{n^{gen}_{MC}(k,l)}$. The
 total statistical error of $c_{_{MC}}$ was calculated as follows:
\begin{equation}\label{eq:deltaCorr}
\Delta c_{_{MC}}=\sqrt{\left(\frac{\Delta\beta}
{\alpha}\right)^2+\left(-\frac{\beta\Delta\alpha}{\alpha^2}\right)^2}~.
\end{equation}
\end{enumerate}

Finally,
the double-differential yield of $\Lambda$ hyperons per inelastic event in
a bin $(k,l)$ amounts to:

\begin{equation}
\label{eq:doubleDiff}
\frac{d^2n}{dkdl}=\frac{c_{dE/dx}c_{_{MC}}(k,l)}{\Delta k \Delta l}
\frac{n^I(k,l)-Bn^R(k,l)}{N^I-BN^R}~,
\end{equation}
\vspace*{-0.1cm}
with
\begin{itemize}
\item $n^{I/R}$ the uncorrected number of $\Lambda$ hyperons for the hydrogen
inserted/removed target configurations,
\item $N^{I/R}$ the number of events for the hydrogen inserted/removed data
after event cuts,
\item $c_{dE/dx}$, $c_{_{MC}}$ the correction factors described in
Sec.~\ref{subs:correction},
\item $B$ the normalisation factor (defined in Sec. \ref{subs:correction}),
\item $k=y$ or $x_{_F}$, and $l=p_{_T}$ or $m_{_T}-m_{_\Lambda}$,
\item $\Delta k$ and $\Delta l$ the bin widths.
\end{itemize}

\subsection{Statistical and systematic uncertainties}
\label{subs:statiAndSyst}
The statistical errors of the corrected double differential yields 
(see Eq.~\ref{eq:doubleDiff}) take into account
the statistical errors of $c_{_{MC}}$ (see Eq.~(\ref{eq:deltaCorr})) and
the statistical errors on the fitted $\Lambda$ yields in the LH inserted and removed
configurations. The statistical errors on $B$ and $c_{dE/dx}$ were neglected.


\label{subs:systematics}
The systematic uncertainties were estimated taking into account four sources. For each source 
modifications to the~standard analysis procedure were applied and the deviation of the results
from the~standard procedure were calculated. As the effects of the modifications 
are partially correlated, the maximal positive and negative deviation from the~standard procedure 
was determined for each bin and source separately. 
Then, the~positive (negative) systematic uncertainties were calculated separately 
by adding in quadrature the~positive (negative) contribution from each source.


The considered sources of the systematic uncertainty and the corresponding modifications of 
the standard method were the following:
\begin{enumerate}[(i)]
\item The uncertainty due to the signal extraction procedure:
\begin{itemize}
\item The standard function used for background fit, a Chebyshev polynomial of 2$^{nd}$ order, 
was changed for a Chebyshev polynomial of 3$^{rd}$ order and for a standard polynomial of 2$^{nd}$ order.
\item The range within which the raw number of $\Lambda$ particles is summed
up was changed from 3$\Gamma$ to 2.5$\Gamma$ and $3.5\Gamma$.
\item The lower limit of the fitting range was  changed from 1.08~\GeVcc (1.076 for $y=0.5$, 1.073 for $y=1.0$) to
1.083~\GeVcc (1.079 for $y=0.5$, 1.076 for $y=1.0$).
\item The initial value of the $\Gamma$ parameter of the signal 
function was changed by $\pm$8~\%.
\item the initial value for the mass parameter of the Lorentz function 
was changed by $\pm$0.3~MeV.
\end{itemize}
\item The effect of the event and quality cuts were checked by
performing the analysis with the following cuts changed compared
to the values presented in Secs.~\ref{subs:eventCuts} and~\ref{subs:candidateCuts}.
\begin{itemize}
\item The cut on the $z$-position of the interaction vertex was changed from 
$\pm$40~cm  to $\pm$30~cm and $\pm$50~cm with respect to the centre of the target.
\item The window in which  off-time beam particles are not allowed was increased from 1~$\mu$s to
1.5~$\mu$s.

\item The elliptic cut on the impact parameters was reduced by 
a factor of 2: $\sqrt{b_{x}^2+(2b_{y})^{2}}< 1$~cm.
\item The \dedx cut was modified to $\pm$2.8$\sigma$ or 3.2$\sigma$ to estimate possible
systematic effects of \dedx calibration.
\item The matching procedure used to reject background in the simulation was turned off.
\item The required minimal number of charge clusters in at least one of the VTPCs for 
both $V^0$-decay products was decreased to 12 or increased to 18.
\item The cut on $\Delta z$, the distance between the decay and the primary
interaction vertex, was changed from the standard values to the values shown in columns A and B 
in the following table: 
\newline


\begin{tabular}{|r|r||c|r@{.}l|r@{.}l|}\hline
\multicolumn{7}{|c|}{Minimal $\Delta z$ (cm) allowed}\\\hline
$y_{min}$&$y_{max}$&standard&\multicolumn{2}{c|}{A}&\multicolumn{2}{c|}{B}\\\hline
-1.75&0.25 &10 &7 &5 &12 &5\\
0.25 &0.75 &15 &11 &25 &18 &75\\
0.75&1.25&40&\multicolumn{1}{r@{\hspace{2pt}}}{30}&&\multicolumn{1}{r@{\hspace{2pt}}}{50}&\\\hline
\end{tabular}
\newline


\item The limits for the cut on $\cos{\phi}$ were changed from the standard values 
to the values shown in columns A and B in the following table: 
\newline

\begin{tabular}{|r|r||r@{.}l|r@{.}l|r@{.}l|}\hline
\multicolumn{8}{|c|}{Maximal $|\cos{\phi}|$ allowed}\\\hline
$y_{min}$&$y_{max}$&\multicolumn{2}{r|}{standard}&\multicolumn{2}{c|}{A}&\multicolumn{2}{c|}{B}\\\hline
-1.75&-0.25&\hspace*{12pt}0&95&0&975&0&925\\
-0.25&0.75&\hspace*{12pt}0&9&0&95&0&85\\
0.75&1.25&\hspace*{12pt}0&8&0&85&0&75\\\hline
\end{tabular}
\newline

\end{itemize}

\item In order to find the systematic uncertainty of the normalisation factor $B$ 
in Eq.~(\ref{eq:B}) for the LH removed configuration,
the limits of the region for which this parameter was calculated was varied 
in steps of 0.1~m. For each combination of the lower limit (ranging from
0.8 to 1.8~m from the target) and upper limit in $z$
(from 2.8~m to 3.8 from the target) the $B$-factor was calculated.
The smallest and the highest value of $B$ obtained in this way is taken as the systematic
uncertainty range of $B$.
\item For estimation of the uncertainty due to the feed-down correction a conservative systematic uncertainty of 30\% on the $\Xi^{-}$ and $\Xi^{0}$ yields predicted by \EposLong was assumed.


\end{enumerate}

The systematic uncertainties are shown in the figures as light blue shaded bars.
They are asymmetric (larger downward) mainly due to the differences
between the results with or without track matching and the change of the background function to
a Chebyshev polynomial of $3^{rd}$ order. For both changes the shift of the results
increases with rapidity. 

\begin{figure}
\begin{center}
\resizebox{0.47\textwidth}{!}{
  \includegraphics{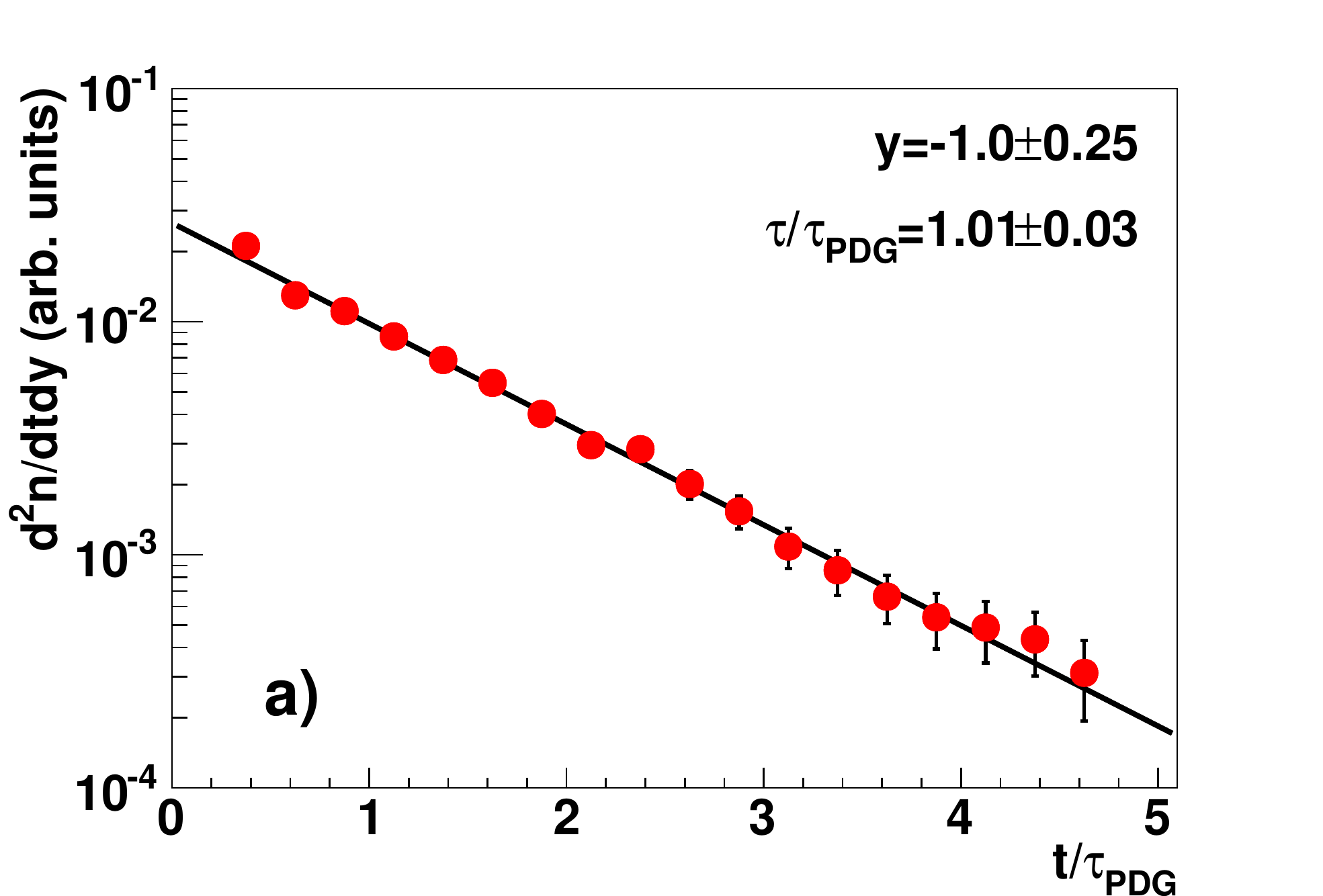}
}
\resizebox{0.47\textwidth}{!}{
  \includegraphics{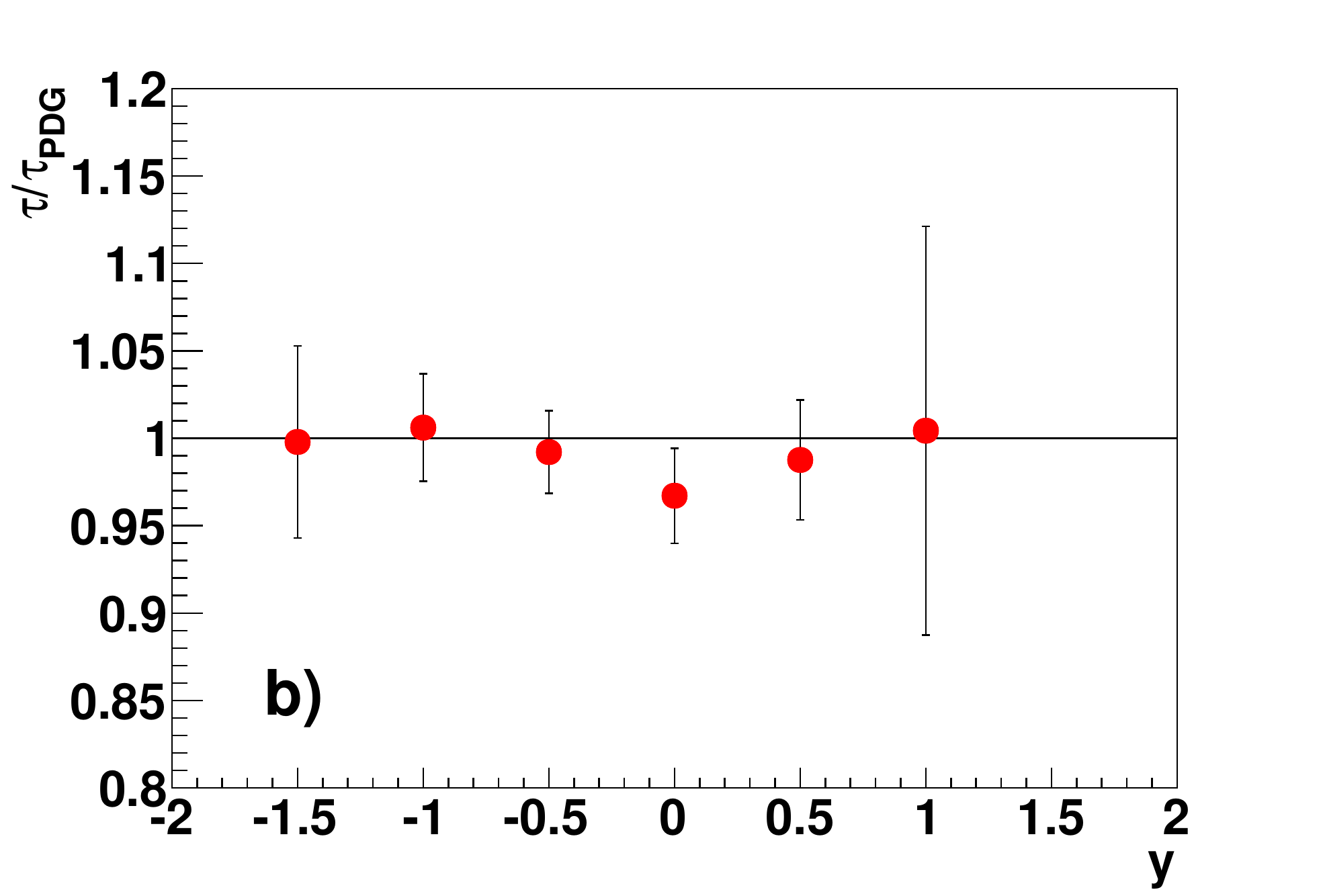}
}
\end{center}
\caption{
{\it Top:} An example of the corrected proper life-time distribution 
for $\Lambda$ hyperons produced in inelastic p+p interactions
at 158~\GeVc in the rapidity interval y=-1.0$\pm$0.25.
{\it Bottom:}
The ratio of the fitted mean life-time to its PDG~\cite{PDG} value as
a function of rapidity.
}
\label{fig:lifetime} 
\end{figure}

The distribution of the proper life-time of  $\Lambda$ hyperons was
obtained using an analysis procedure analogous to the one presented in Sec.~\ref{sec:results}.
The data for the lifetime analysis were binned in rapidity $k=y$ (from -1.5 to
+1.0, in steps of 0.5) and life-time normalised to the mean lifetime
$t/\tau_{PDG}$~\cite{PDG} (from 0.00 to 4.75, in steps of 0.25) with
$c\tau_{PDG}=7.89$~cm. The life-time was calculated using the distance $r$
between the V$^0$-decay vertex and the interaction vertex of
the V$^0$-candidates ($t=r/(\gamma\beta)$, where
$\gamma$, $\beta$ are the Lorentz variables). 
Then $d^2n/(dydt)$ was calculated and an exponential function was fitted 
to the life-time distribution for each rapidity bin separately
(see the example in Fig.~\ref{fig:lifetime}~(a) for $y=-1.0$).
The ratio of the fitted mean life-time $\tau$ to the corresponding PDG
value $\tau_{PDF}$ is shown in  Fig.~\ref{fig:lifetime}~(b) as a function of rapidity.
The fitted mean life-times are seen to agree with the PDG value for
all rapidities indicating good accuracy of the correction procedure.

The expected forward-backward symmetry of the data was also checked.
The final double- and single-differential
distributions used for this test were found to agree for the corresponding
backward and forward rapidities within the statistical errors.

In addition, the stability of results in different periods during the data taking was investigated. For that purpose, the data set was divided into two subsets, containing runs from the first and the second half of the data taking period. These subsets were analysed separately and the results are found to be consistent.

\section{Results}
\label{sec:results}

\subsection{Formalism}

The double-differential yields of $\Lambda$ hyperons in inelastic p+p interactions at 158 GeV/$c$
were calculated in kinematic $(k,l)$ bins (with $k=y$ or $x_{_F}$, and $l=p_{_T}$ or $m_{_T}-m_{_\Lambda}$) 
using Eq.~(\ref{eq:doubleDiff}). The following spectra are presented: $\frac{d^2n}{dydp_{_T}}$,
$\frac{d^2n}{dydm_{_T}}$, $\frac{d^{2}n}{dx_{F}dp_{T}}$ and $f_n(x_{_F},p_{_T})$, where
\begin{equation}
  \label{eq:f}
  \begin{split}
    f_n(x_{_F},p_{_T})&= \frac{2E^*}{\pi\sqrt{s}}\frac{d^2n}{dx_{_F}dp^2_{_T}}=
  \frac{1}{\pi\sqrt{s}}\frac{E^*}{p_{_T}}\frac{d^2n}{dx_{_F}dp_{_T}}~.
  \end{split}
\end{equation}
$E^{*}$ is the energy of the $\Lambda$ hyperon in the centre of mass
system. 
The weighting factor, $E^*/p_{_T}$, was calculated at the centre of each $(x_{_F},p_{_T})$ bin and is consistent with the average value $\left\langle E^*/p_{_T}\right\rangle$ obtained using the EPOS generator.


Single-differential $\frac{dn}{dk}$ distributions are obtained by summing
the double-differential yields for a given $k$ over $l$. In order to estimate
the yield in the unmeasured high $p_{T}$ range, the function
\begin{equation}
\label{eq:scaling}
u =
\frac{1}{p_{_T}}\frac{d^2n}{dkdp_{_T}}=\frac{1}{m_{_T}}\frac{d^2n}{dkdm_{_T}}=Ae^{-\frac{m_{_T}}{T}}
\end{equation}
was fitted to the data and integrated beyond the measured $p_{T}$, where
$A$ denotes the normalisation factor and $T$ the
inverse slope parameter. Single-differential invariant yields $F_n(x_{_F})$
were obtained by performing an integration of Eq.~(\ref{eq:f}) with respect
to $p_{_T}^2$:
\begin{equation}
\label{eq:F}
F_n(x_{_F})=\int\limits_0^{+\infty}f_n(x_{_F},p_{_T})dp_{_T}^2=
\frac{2}{\pi\sqrt{s}}\int\limits_0^{+\infty}{E^*
\frac{d^2n}{dx_{_F}dp_{_T}}dp_{_T}}~.
\end{equation}
For the calculation of $F_n(x_{_F})$,
the weighting factor $E^*$ was calculated at the centre of each $(x_{_F},p_{_T})$ bin. For the extrapolation into the unmeasured high $p_{T}$ region
Eq.~(\ref{eq:scaling}) was used.

Invariant cross-section is obtained from $F_n(x_{_F})$ by multiplying it by the inelastic cross-section $\sigma_{inel}$:
\begin{equation}
\label{eq:Fsigma}
F(x_{_F})=\sigma_{inel}F_n(x_{_F})~.
\end{equation}

The mean transverse mass $\langle m_{_T}\rangle$ was calculated
from the function Eq.~(\ref{eq:scaling}) fitted to the $m_{_T}$
distribution as follows:
\begin{equation}
\label{eq:meanmT}
\langle
m_{_T}\rangle=\frac{\int_0^{+\infty}{m_{_T}u(m_{_T})dm_{_T}}}{\int_0^{+\infty}{u(m_{_T})dm_{_T}}}.
\end{equation}


\subsection{Spectra}
\label{subs:spectra}

\begin{figure*}
\begin{center}
\resizebox{0.99\textwidth}{!}{
  \includegraphics{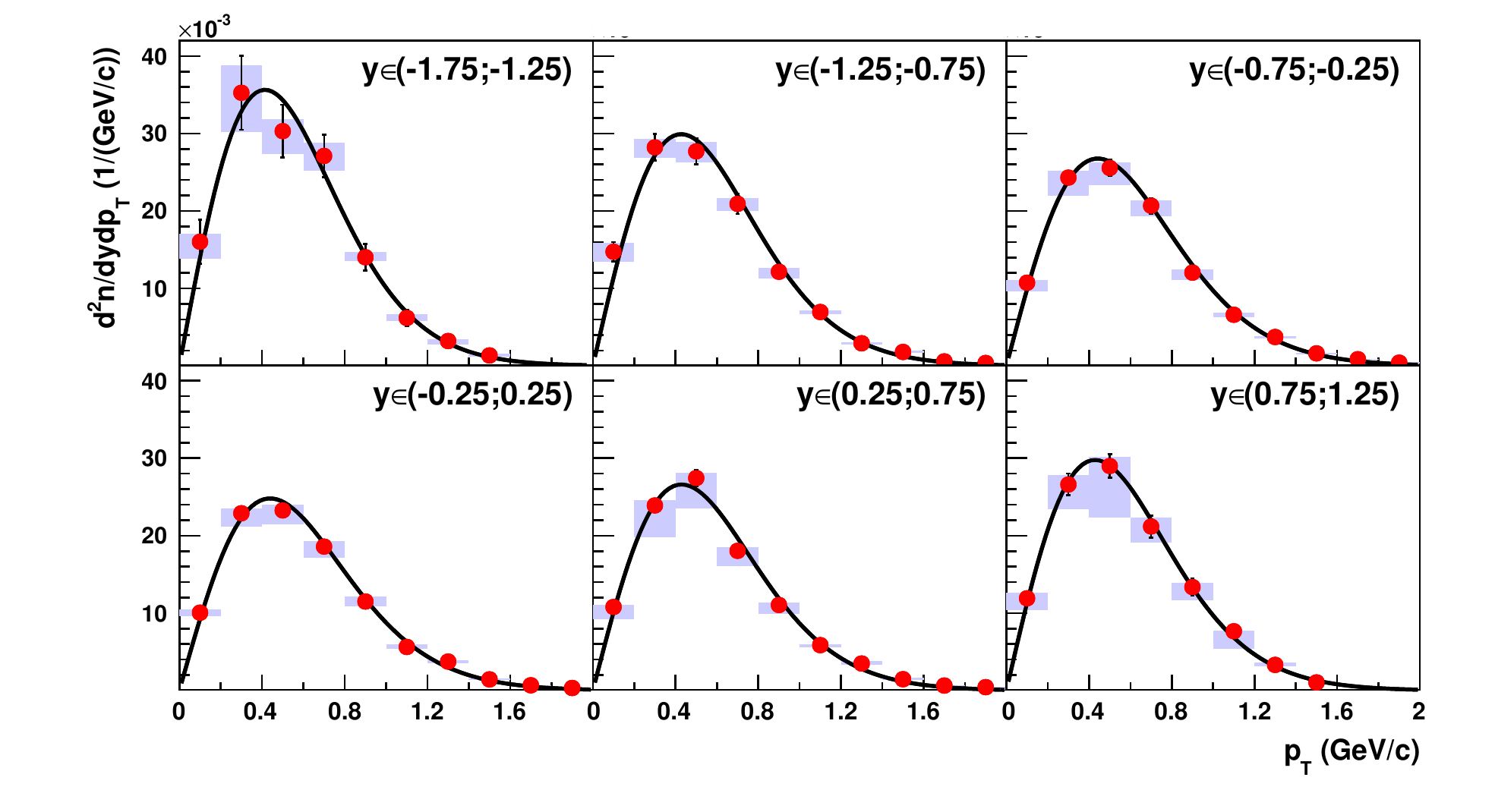}
}
\end{center}
\caption{Spectra $\frac{d^2n}{dydp_{_T}}$ for six bins in rapidity
$y$. The fitted function is given by Eq.~(\ref{eq:scaling}).
The numerical data are listed in the Table~\ref{tab:d2ndydpT} and  
the fitted inverse slope parameter for each of the bins in Table~\ref{tab:dndy}.} 
\label{fig:pT} 
\end{figure*}

\begin{figure*}
\begin{center}
\resizebox{0.99\textwidth}{!}{
  \includegraphics{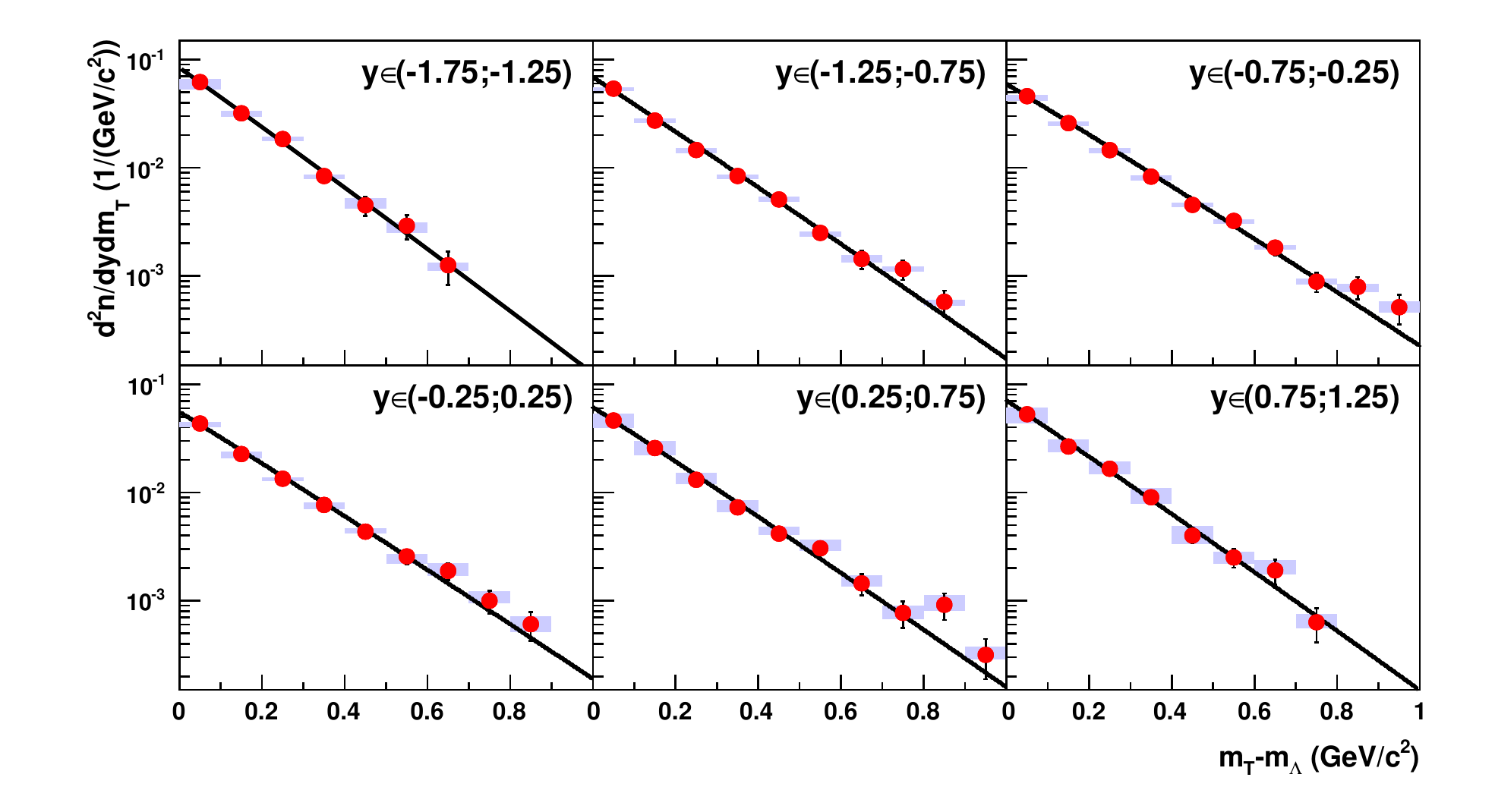}
}
\end{center}
\caption{Spectra $\frac{d^2n}{dydm_{_T}}$ for six bins in rapidity
$y$. The fitted function is given by Eq.~(\ref{eq:scaling}).
The numerical data are listed in Table~\ref{tab:d2ndydmT} and the mean
transverse mass $\langle m_{_T}\rangle -m_{_\Lambda}$ for each of the bins in
Table~\ref{tab:dndy}.
} 
\label{fig:mT} 
\end{figure*}

\begin{figure*}
\begin{center}
\resizebox{0.99\textwidth}{!}{
  \includegraphics{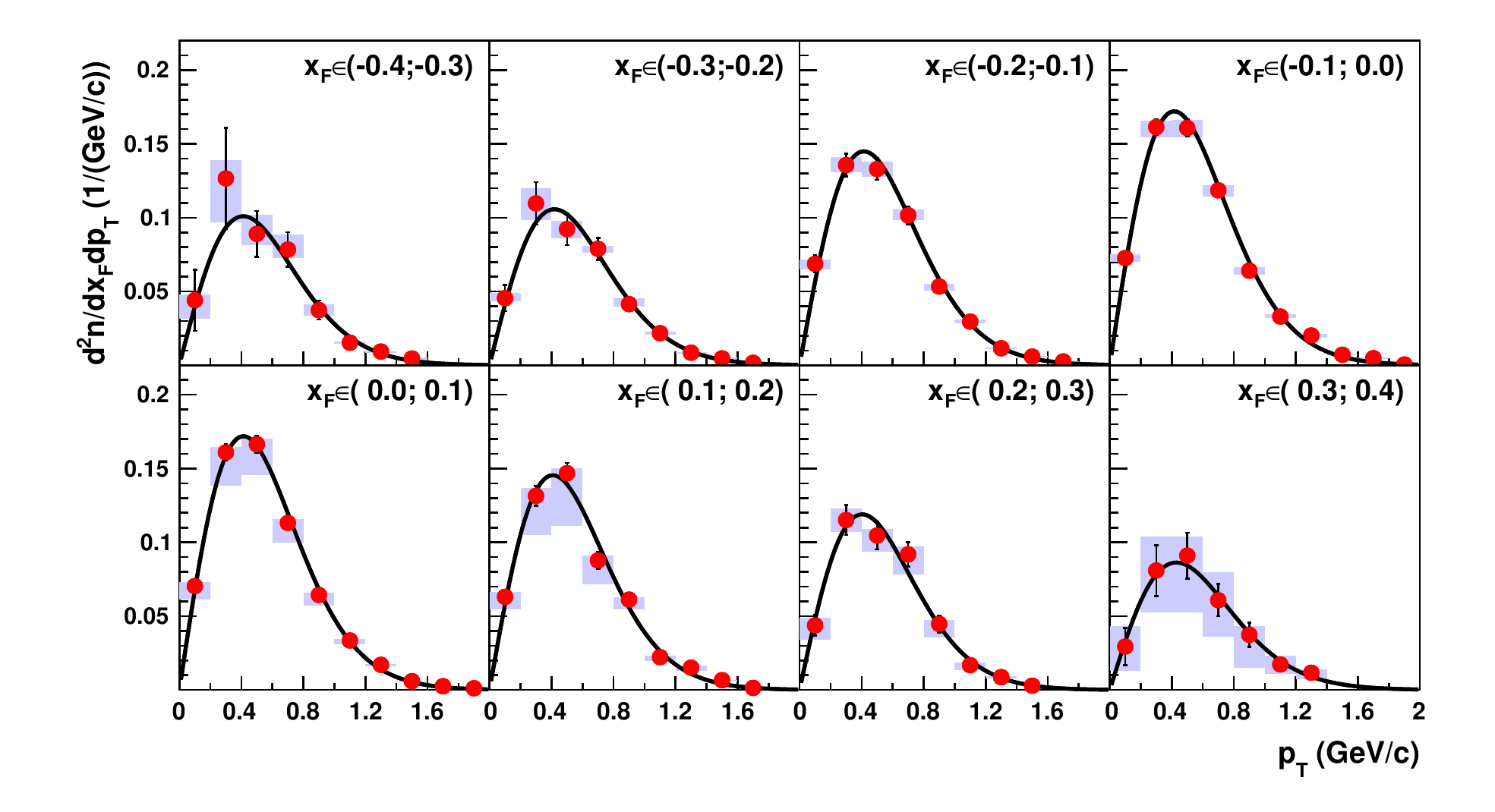}
}
\end{center}
\caption{Spectra $\frac{d^2n}{dx_{_F}dp_{_T}}$ for eight
bins in $x_{_F}$. The fitted function is given by Eq.~(\ref{eq:scaling}).
The numerical data are listed in Table~\ref{tab:d2ndxFdpT} for
$x_{_F} > 0$ and in Table~\ref{tab:2d2ndxFdpT} for $x_{_F} > 0$.
}
\label{fig:xF} 
\end{figure*}

\begin{figure*}
\begin{center}
\resizebox{0.99\textwidth}{!}{
  \includegraphics{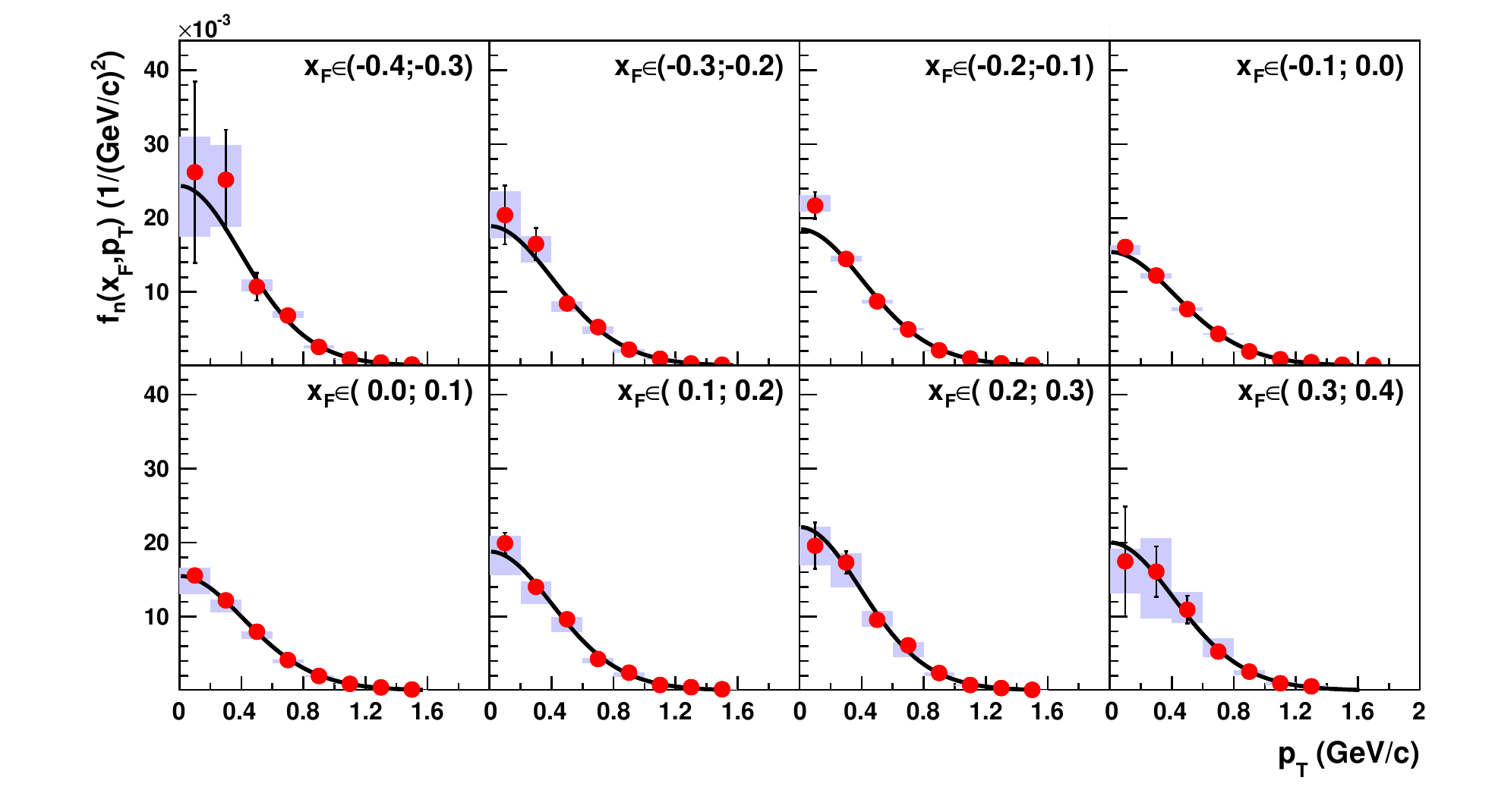}
}
\end{center}
\caption{Spectra $f_n$ for eight bins in $x_{_F}$.
The fitted function was obtained from Eq.~(\ref{eq:scaling})
by multiplying the right-hand side by $p_{_T}$. The numerical data are listed in Table~\ref{tab:d2ndxFdpT}
for $x_{_F} > 0 $ and in Table~\ref{tab:2d2ndxFdpT} for $x_{_F} < 0$.}
\label{fig:xFRew} 
\end{figure*}

Double-differential $\frac{d^2n}{dydp_{_T}}$, $\frac{d^2n}{dydm_{_T}}$, 
$\frac{d^{2}n}{dx_{F}dp_{T}}$ and $f_n(x_{_F},p_{_T})$ spectra are shown in
Figs.~\ref{fig:pT} - \ref{fig:xFRew}.  The numerical values for
$\frac{d^2n}{dydp_{_T}}$ and $\frac{d^2n}{dydm_{_T}}$ are presented in 
Tables~\ref{tab:d2ndydpT} and~\ref{tab:d2ndydmT}, 
while invariant and non-invariant $x_{F}$ yields are shown in 
Table~\ref{tab:d2ndxFdpT} for $x_{_F}<0$ and
Table~\ref{tab:2d2ndxFdpT} for $x_{_F}>0$.

The values of the $p_{T}$ integrated $\frac{dn}{dy}$ rapidity distribution
are presented in Table~\ref{tab:dndy}.  The table also contains the values
of the inverse slope parameter T and the mean transverse mass $\langle
m_{_T}\rangle -m_{_\Lambda}$ as function of rapidity. 
The single-differential $x_{F}$ distributions
are summarised in Table~\ref{tab:dndxF}. In Sec.~\ref{sec:comparison} below,
the obtained single-differential distributions are compared to model
predictions and previously published experimental results.

The mean multiplicity of $\Lambda$ hyperons ($\langle\Lambda\rangle$) was
determined from the $x_{_F}$ distribution.  As the models applicable
in the SPS energies range show large discrepancies in the region not measured by \NASixtyOne
(see Fig.~\ref{fig:mc}), the $\Lambda$ yield in the unmeasured $x_{F}$ region
($|x_{F}|$ > 0.4) was approximated by the straight line shown in Fig.~\ref{fig:mc}.
The line is defined assuming symmetry of the distribution. It crosses the points
$A_\pm=\left(\pm0.35,\frac{1}{2}\left(\frac{dn}{dx_{_F}}(-0.35)+\frac{dn}{dx_{_F}}(0.35)\right)\right)$
and $B_\pm=(\pm1,0)$. For the estimation of statistical part of the extrapolation error, 
the value of the point A was increased/decreased by
$\frac{1}{2}\left(\Delta\frac{dn}{dx_{_F}}(-0.35)+\Delta\frac{dn}{dx_{_F}}(0.35)\right)$.
The extrapolation amounts to 34.3\% of the total $\Lambda$ yield and results in
$\langle\Lambda\rangle=0.120\pm0.006\;(stat.)$ of the mean $\Lambda$ multiplicity.

For the \Epos model, not used for this extrapolation, 
the yield outside of \NASixtyOne acceptance to the total yield amounts to 38.0~\%.

The systematic uncertainty of the mean multiplicity was calculated following the procedure described
in Sec \ref{subs:statiAndSyst}. An additional source of systematic uncertainty 
arises from the extrapolation of the $\Lambda$ multiplicity to full phase-space.
This was estimated by an alternative procedure based on a parametrisation of
published rapidity distributions. 
In an iterative procedure a symmetric polynomial of 4$^{th}$ order \cite{Marek2} was fitted
to the $(1/\langle n\rangle)(dn/dz)$ distributions obtained by five bubble-chamber
experiments~\cite{Ammosov,Chapman,Brick,Jaeger,LoPinto} and the \NASixtyOne data, where $z$ stands
for $y/y_{beam}$. First, the fit included only the five bubble-chamber datasets. 
Next, the \NASixtyOne spectrum was normalised to the fit result
obtained in the first step and added as the 6$^{th}$ set for
the fit. Finally, the procedure was iterated using those six datasets until
the normalisation factor converged.
The ratio of the integral of the fitted function
$\frac{1}{\langle\Lambda\rangle}\frac{dn}{dz}\left(z\right)=0.394+1.99z^2-2.66z^4$
(see Fig.~\ref{fig:world}) for the full range of rapidity to the integral in
the range outside of the \NASixtyOne acceptance was used as the extrapolation factor 
for the \NASixtyOne results. This ratio amounted to $1.92\pm0.12$, i.e. 48~\% of
the total production is outside of the acceptance for this procedure
resulting in a mean multiplicity of $\langle\Lambda\rangle=0.129\pm0.008$. 
The difference between this result and the linear extrapolation of the $x_F$ distribution
is added in quadrature to the (positive) systematic error.

The final result for the $\Lambda$ multiplicity in inelastic
p+p interactions at 158 GeV/$c$ then reads as follows:
$$\langle\Lambda\rangle=0.120\pm0.006\;(stat.)\;\pm0.010\;(sys.)$$

\section{Comparison with world data and model predictions}
\label{sec:comparison}

\begin{figure}
\begin{center}
\resizebox{0.47\textwidth}{!}{
  \includegraphics{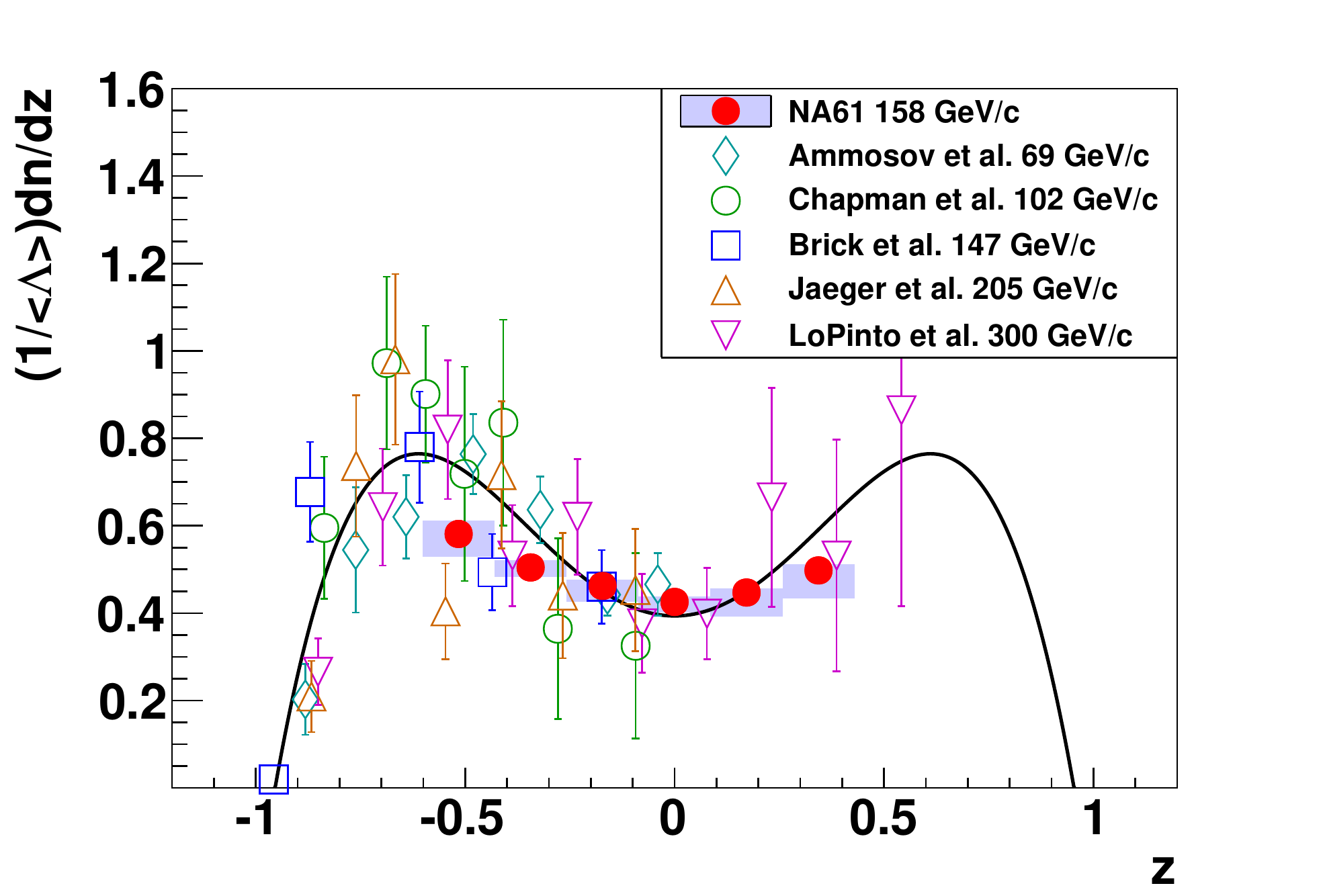}
}
\end{center}
\caption{The scaled $\Lambda$ yield as function of scaled rapidity $z=y/y_{beam}$
 in inelastic p+p interactions measured by 
\NASixtyOne  and selected bubble-chamber
experiments~\cite{Ammosov,Chapman,Brick,Jaeger,LoPinto}.
The symmetric polynomial of 4$^{th}$ order used for estimation of the
systematic uncertainty of $\Lambda$ mean multiplicity
(see Sec.~\ref{subs:spectra}) is plotted to guide the eye.
}
\label{fig:world} 
\end{figure}

The single-differential spectra from \NASixtyOne are compared in Fig.~\ref{fig:world} to
results from five bubble-chamber experiments which measured p+p interactions
at beam momenta close to 158~\GeVc.
The experiments published data for the backward hemisphere,
however, with rather small statistics~\cite{Ammosov,Chapman,Brick,Jaeger,LoPinto}
and correspondingly large uncertainties. 
In order to account for the difference in beam momentum
the spectra are shown in terms of the scaled rapidity $z=y/y_{beam}$
and were normalised to unity in order to compare the shapes.
Note, that the same data sets were also used to compute the alternative
correction factor used to estimate the systematic uncertainty
of $\langle\Lambda\rangle$ (see Sec.~\ref{sec:results})
obtained by \NASixtyOne.

Though the statistical error and the systematic uncertainty of the \NASixtyOne
measurement is much smaller than
for the other experiments, and the results are consistent with all the
datasets used for the comparison, the general tendency obtained
by fitting a symmetric polynomial of 4$^{th}$ order does not describe well
the \NASixtyOne data. On the other hand, the result of Brick {\it et al.}
for which the beam momentum (147~\GeVc) differs the least from the
\NASixtyOne momentum, shows the best agreement.

\begin{figure}
\begin{center}
\resizebox{0.47\textwidth}{!}{
  \includegraphics{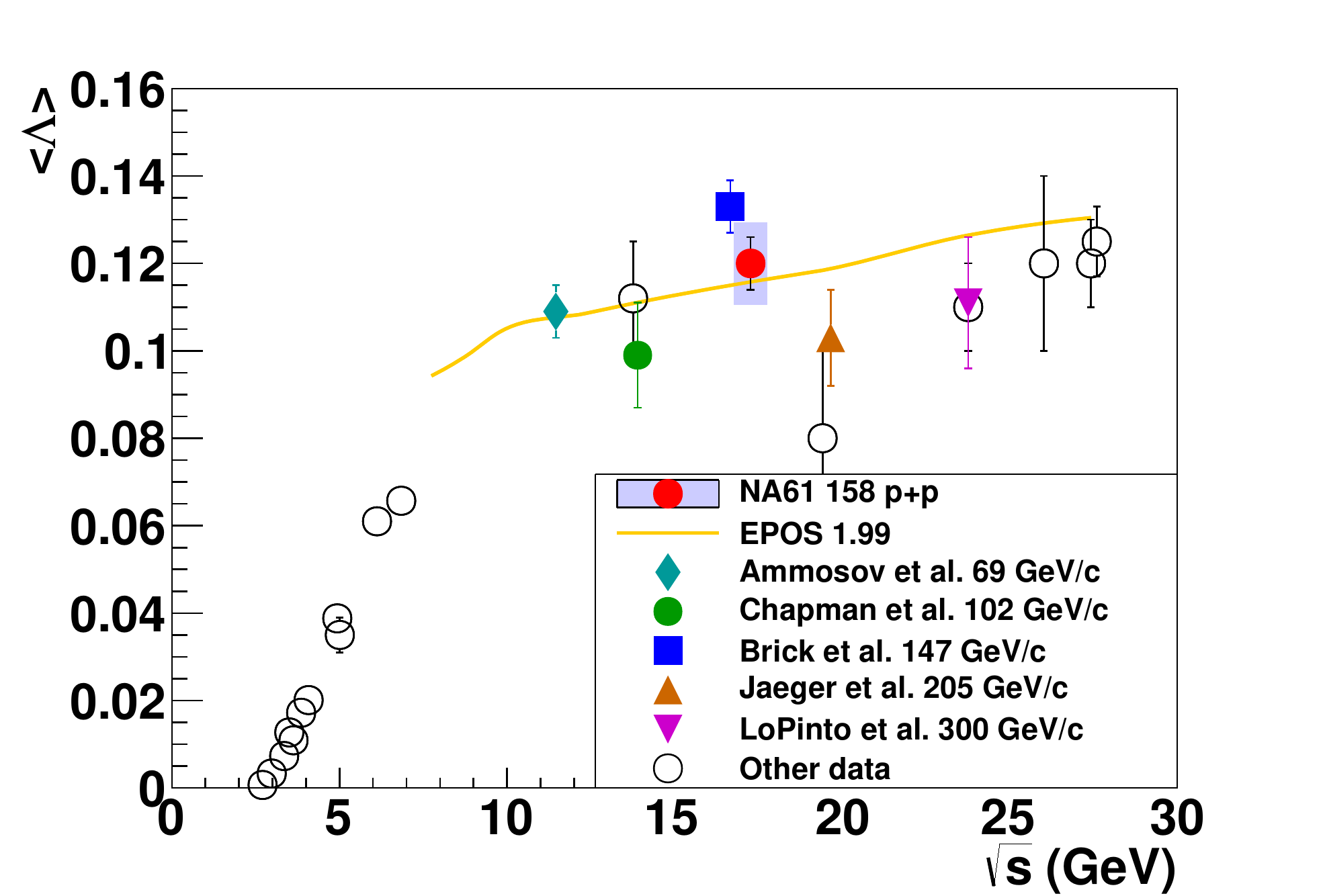}
}
\end{center}
\caption{Collision energy dependence of mean multiplicity of $\Lambda$ 
hyperons produced in inelastic p+p interactions.
Full symbols indicate
bubble chamber results~\cite{Ammosov,Chapman,Brick,Jaeger,LoPinto},
the solid red dot shows the \NASixtyOne result.
Open symbols depict the remaining world data~\cite{Marek}.
The \EposLong~\cite{EPOS} prediction is shown by the curve. The
systematic uncertainty of the \NASixtyOne result is indicated
by the shaded bar.}
\label{fig:meanMulti} 
\end{figure}

The mean multiplicity of $\Lambda$ for 158~\GeVc inelastic p+p interactions
is compared in Fig.~\ref{fig:meanMulti} with the world data~\cite{Marek} 
as well as with predictions of the \EposLong model in its validity range.
A steep rise in the threshold region is followed by a more gentle increase
at higher energies that is well reproduced by the \EposLong model.

\begin{figure}
\begin{center}
\resizebox{0.47\textwidth}{!}{
  \includegraphics{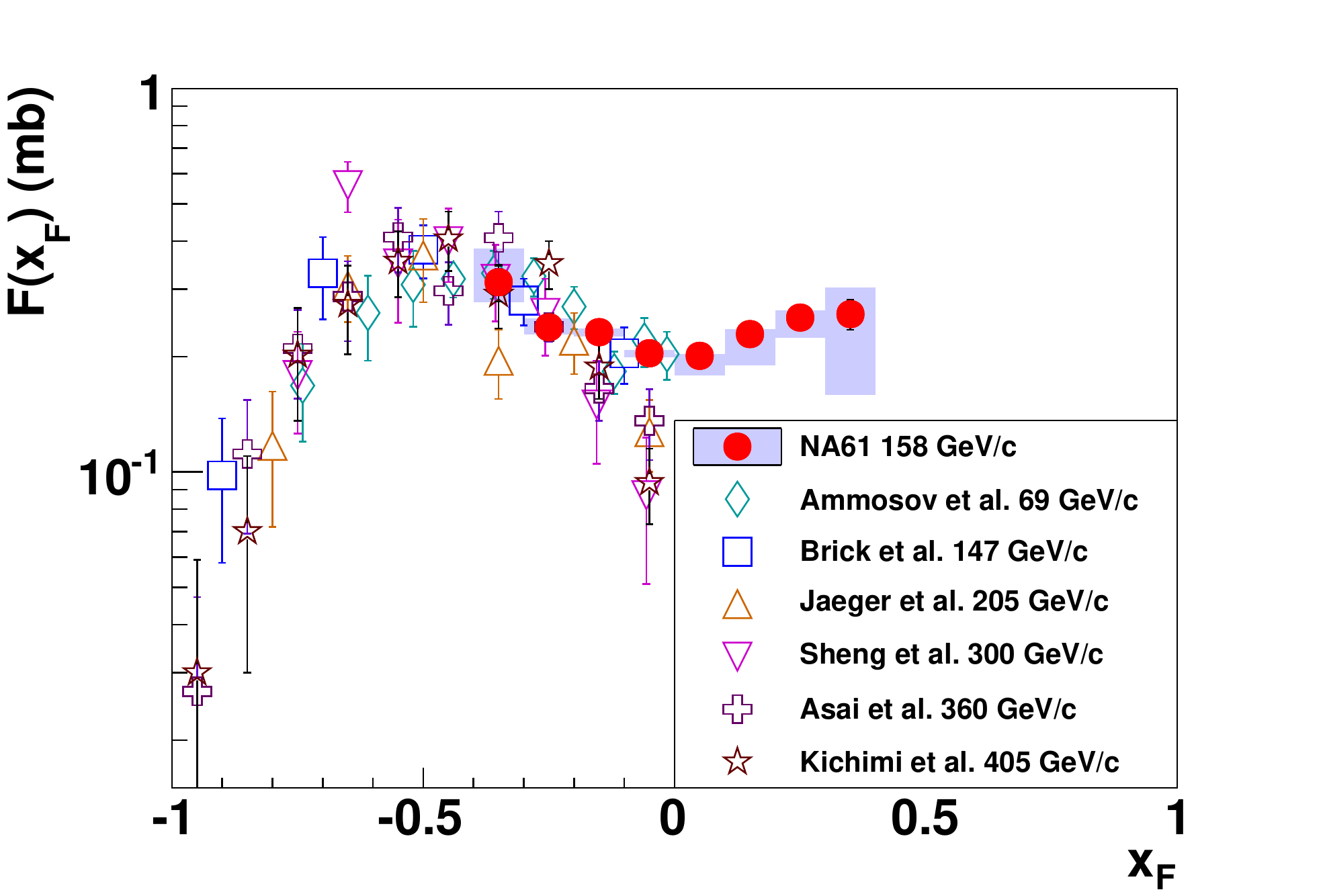}
}

\end{center}
\caption{
  Comparison of the invariant $x_{F}$ spectra (see Eq.~(\ref{eq:Fsigma}), 
  $\sigma_{inel.}=31.8$~mb~\cite{PDG}) from the \NASixtyOne with the data from bubble-chamber experiments at beam
  momenta close to 158~\GeVc~\cite{Ammosov,Brick,Jaeger,Sheng,Asai,Kichimi}.
  The \NASixtyOne data points are indicated by filled circles. 
  The systematic uncertainty is shown as a band around the points. 
}
\label{fig:worldxF} 

\end{figure}

The dependence of the invariant spectrum on $x_{_F}$ for \NASixtyOne and 
published results from bubble chamber experiments~\cite{Ammosov,Brick,Jaeger,Sheng,Asai,Kichimi}
at nearby beam momenta is shown in Fig.~\ref{fig:worldxF}. The \NASixtyOne 
results are consistent with the experiments performed at
proton beams of lower energy, although the dip-like structure visible 
at central $x_{_F}$ in the data from the experiments
operating at higher energies is not observed. 

\begin{figure}
\begin{center}
\resizebox{0.47\textwidth}{!}{
  \includegraphics{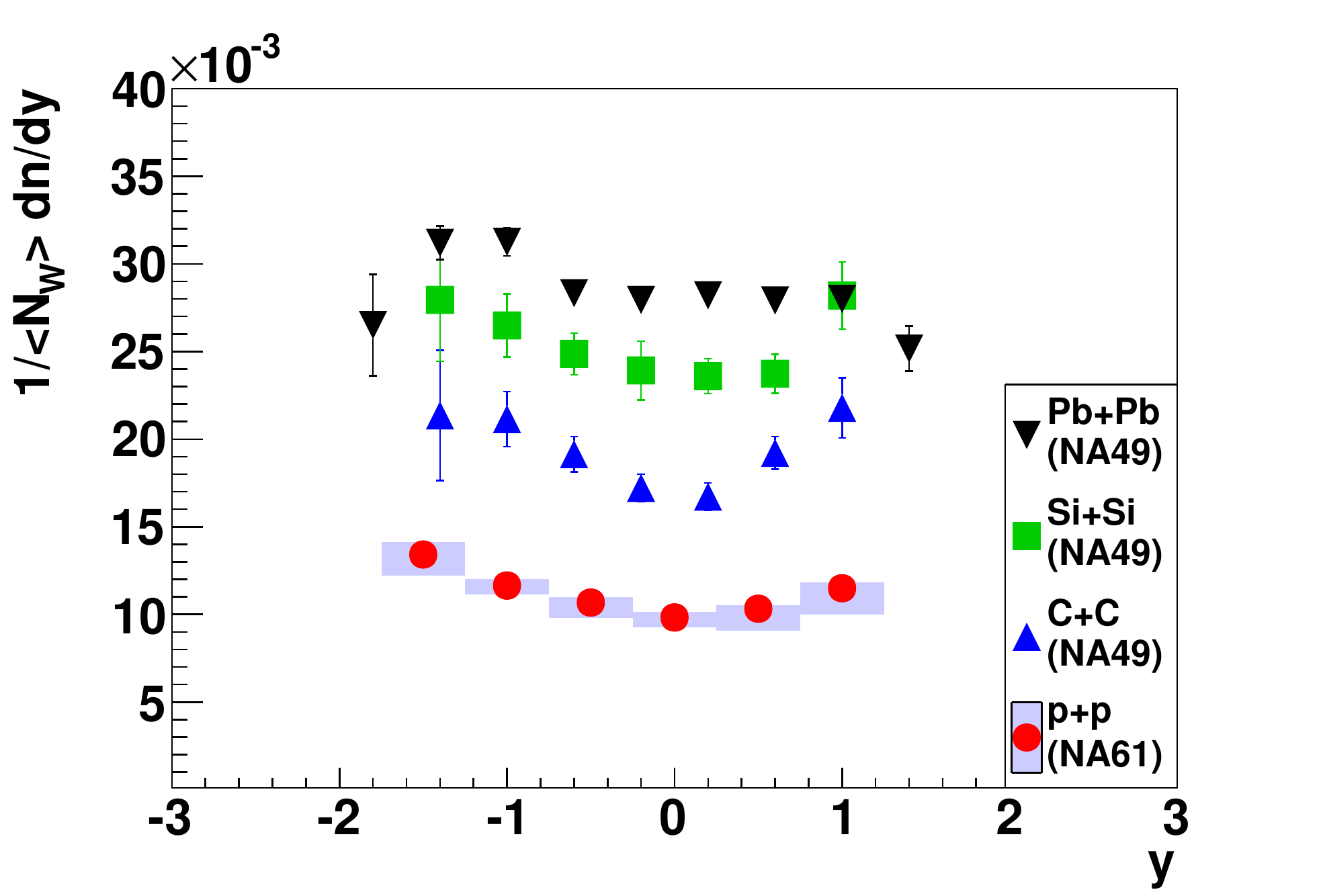}
}
\end{center}
\caption{Rapidity spectra  of $\Lambda$ hyperons divided by
the mean number of wounded nucleons in inelastic p+p interactions (\NASixtyOne) 
and central C+C, Si+Si~\cite{CCSiSi}, and Pb+Pb~\cite{PbPb} collisions
at 158~\AGeVc. }
\label{fig:na49comp} 
\end{figure}

Figure~\ref{fig:na49comp} shows a comparison of rapidity spectra
divided by the mean number of wounded nucleons $1/\langle N_W\rangle$
in inelastic p+p interactions (this paper) and central C+C, Si+Si and
Pb+Pb collisions (NA49~\cite{CCSiSi,PbPb}) at 158~\AGeVc. 
The yield of $\Lambda$ hyperons per wounded nucleon increases with 
increasing $\langle N_W\rangle$ as a consequence of strangeness enhancement
in nucleus-nucleus collisions.

\begin{figure}
\begin{center}
\resizebox{0.47\textwidth}{!}{
  \includegraphics{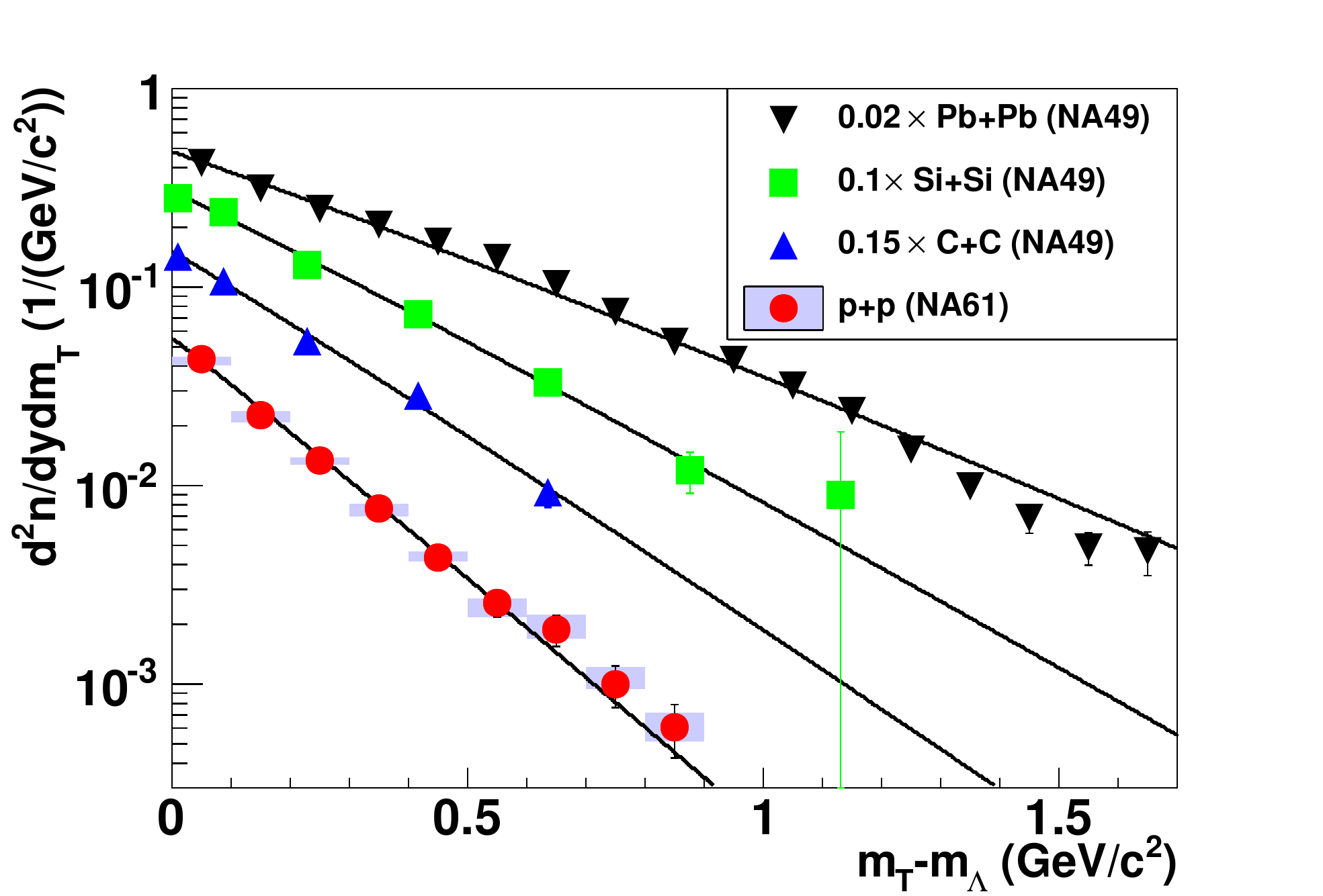}
}
\end{center}
\caption{Spectra of m$_{_T}$ at mid-rapidity ($|y|\leq0.4$ for A+A,
$|y|\leq0.25$ for p+p)
from \NASixtyOne inelastic p+p interactions and central C+C, Si+Si~\cite{CCSiSi}, and
Pb+Pb~\cite{PbPb} collisions for beam momentum of 158~\AGeVc. The lines are fitted using
Eq.~(\ref{eq:scaling}).}
\label{fig:na49compmT} 
\end{figure}

Figure~\ref{fig:na49compmT} displays $m_T$ spectra at mid-rapidity for
inelastic p+p interactions (this paper) and central nucleus-nucleus collisions
(NA49~\cite{CCSiSi,PbPb}) at 158~\AGeVc. 
The inverse slope parameter of the spectrum increases with 
increasing nuclear size due to increasing transverse flow.

\begin{figure}
\begin{center}
\resizebox{0.47\textwidth}{!}{
  \includegraphics{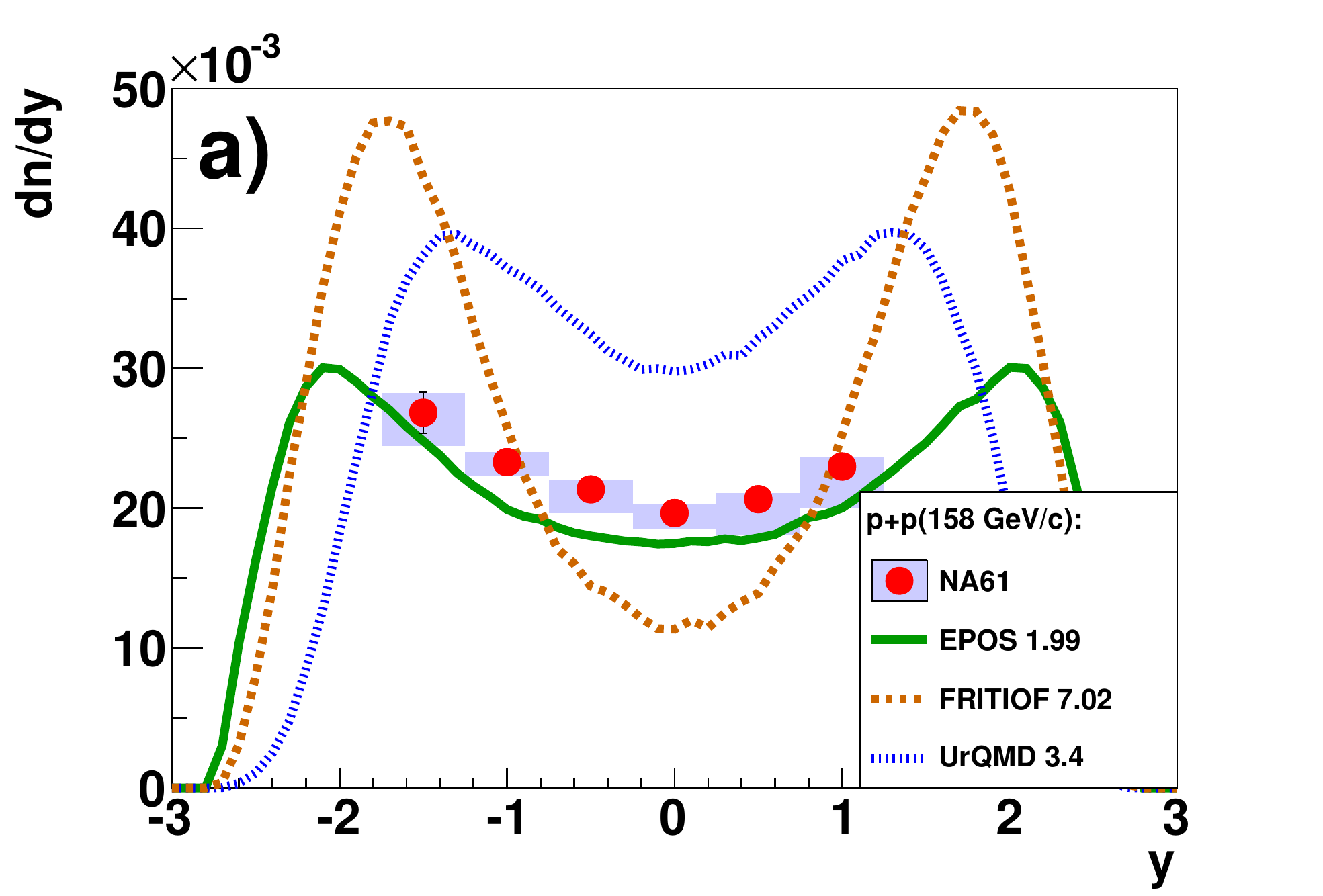}
}
\resizebox{0.47\textwidth}{!}{
  \includegraphics{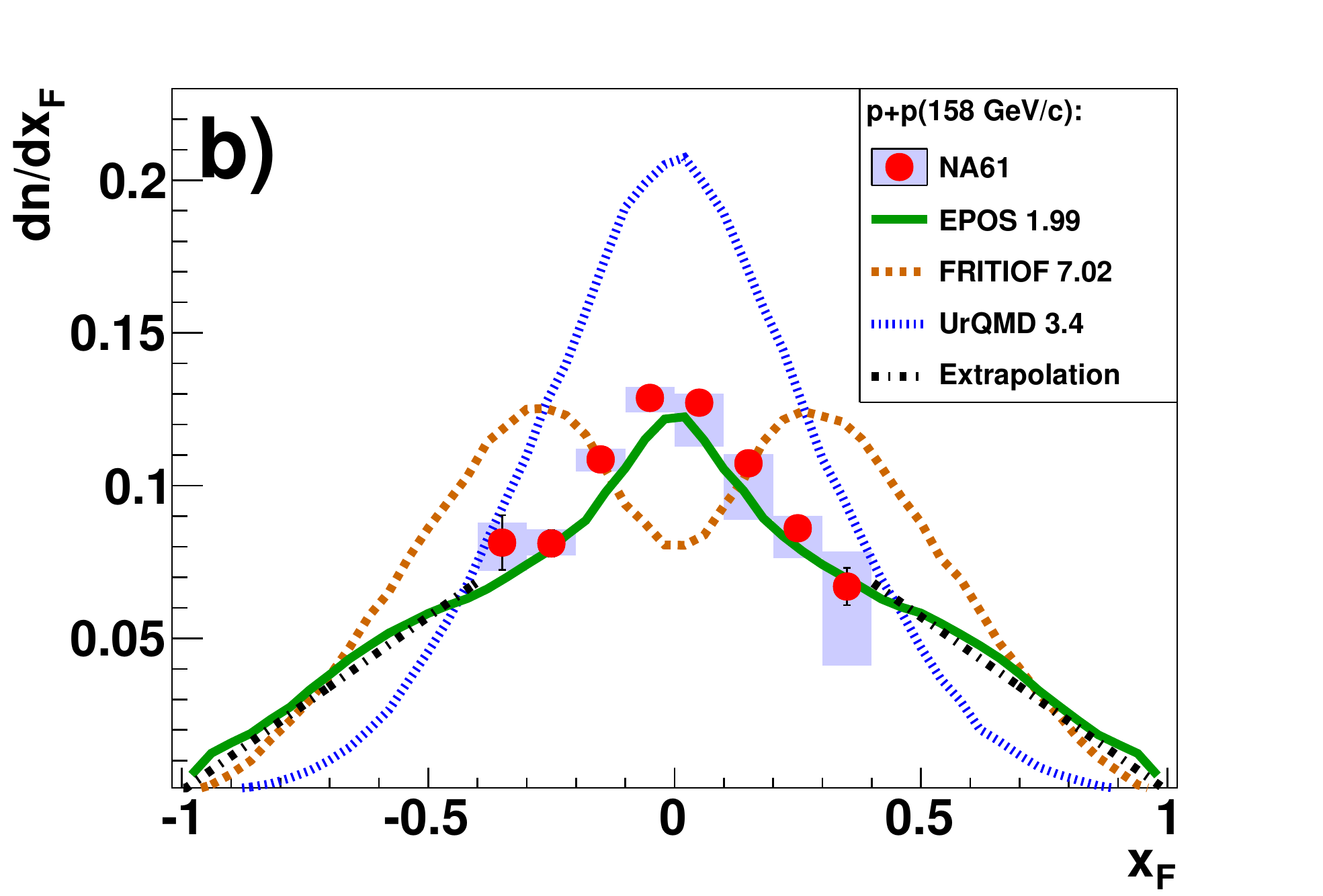}
}

\end{center}
\caption{Comparison of $\frac{dn}{dy}$ (a), and $\frac{dn}{dx_{_F}}$ (b)
distributions with calculations of the \Epos~\cite{EPOS},
\Urqmd~\cite{UrQMD1,UrQMD2} and \Fritiof~\cite{FRITIOF} models.
The chain line was used to extrapolate the \NASixtyOne
measurements to full phase space. For details see text.
}
\label{fig:mc} 
\end{figure}

A comparison with calculations from the models~\EposLong~\cite{EPOS}, 
\UrqmdLong~\cite{UrQMD1,UrQMD2}, and \FritiofLong~\cite{FRITIOF} embedded in \HsdLong~\cite{HSD} is presented  
in Fig.~\ref{fig:mc}.
The best agreement is found for the \EposLong model.

\section{Summary}
Inclusive production of \mbox{$\Lambda$-hyperons} was measured
with the large acceptance \NASixtyOne spectrometer at the CERN SPS
in inelastic p+p interactions at beam momentum of 158~\GeVc.
Spectra of transverse momentum (up to 2~\GeVc) and transverse mass
as well as distributions of rapidity (from -1.75 to 1.25) and x$_{_F}$ (from -0.4 to 0.4)
are presented. The~mean multiplicity was found to be
$0.120\pm0.006\;(stat.)\;\pm0.010\;(sys.)$.
The new results are in reasonable agreement with measurements from
bubble-chamber experiments at nearby beam momenta, but have much smaller
uncertainties.
 
Predictions of the \Epos, \Urqmd and \Fritiof models were
compared to the new \NASixtyOne measurements reported in this paper.
While \Epos describes the data quite well, significant discrepancies 
are observed with the latter two models. 

The results expand our knowledge of elementary proton-proton interactions, allowing for 
a more precise description of strangeness production. They are expected to be used not only as
an important input in the research of strongly interacting matter, but also as an input
for tuning MC-generators, including those used for cosmic-ray shower and neutrino beams simulations.

\section*{Acknowledgements}
This work was supported by
the Hungarian Scientific Research Fund (grants OTKA 68506 and 71989),
the J\'anos Bolyai Research Scholarship of
the Hungarian Academy of Sciences,
the Polish Ministry of Science and Higher Education (grants 667\slash N-CERN\slash2010\slash0, NN\,202\,48\,4339 and NN\,202\,23\,1837),
the Polish National Science Centre (grants~2011\slash03\slash N\slash ST2\slash03691, 2012\slash04\slash M\slash ST2\slash00816 and 
2013\slash11\slash N\slash ST2\slash03879),
the Foundation for Polish Science --- MPD program, co-financed by the European Union within the European Regional Development Fund,
the Federal Agency of Education of the Ministry of Education and Science of the
Russian Federation (SPbSU research grant 11.38.193.2014),
the Russian Academy of Science and the Russian Foundation for Basic Research (grants 08-02-00018, 09-02-00664 and 12-02-91503-CERN),
the Ministry of Education, Culture, Sports, Science and Tech\-no\-lo\-gy, Japan, Grant-in-Aid for Sci\-en\-ti\-fic Research (grants 18071005, 19034011, 19740162, 20740160 and 20039012),
the German Research Foundation (grant GA\,1480/2-2),
the EU-funded Marie Curie Outgoing Fellowship,
Grant PIOF-GA-2013-624803,
the Bulgarian Nuclear Regulatory Agency and the Joint Institute for
Nuclear Research, Dubna (bilateral contract No. 4418-1-15\slash 17),
Ministry of Education and Science of the Republic of Serbia (grant OI171002),
Swiss Nationalfonds Foundation (grant 200020\-117913/1)
and ETH Research Grant TH-01\,07-3.
Finally, it is a pleasure to thank the European
Organisation for Nuclear Research for strong support and
hospitality and, in particular, the operating crews of the CERN
SPS accelerator and beam lines who made the measurements
possible.

\newpage

\begin{table}
\caption{Double-differential yield $\frac{d^2n}{dydp_{_T}}$.} 

\label{tab:d2ndydpT}       
\begin{center}
\begin{tabular}{|r|r@{.}l|r@{.}l r@{.}l r@{.}l r@{.}l|}\hline
\multicolumn{1}{|r|}{$y$}&\multicolumn{2}{|l|}{$p_{_T}$}&\multicolumn{2}{l}{$\frac{d^2n}{dydp_{_T}}$}&\multicolumn{2}{l}{$\Delta_{stat}$}
&\multicolumn{2}{l}{$\Delta_{sys}^-$}&\multicolumn{2}{l|}{$\Delta_{sys}^+$}\\ 
&\multicolumn{2}{l|}{}&\multicolumn{8}{l|}{$\times10^3$ ($\frac{1}{GeV/c}$)}\\\hline 
\multirow{8}{*}{-1.5}             &0&1       &16&0        &2&8     &2&2        &1&0   \\
                                  &0&3       &35&3        &4&8     &5&0        &3&5   \\
                                  &0&5       &30&3        &3&4     &2&9        &1&6   \\
                                  &0&7       &27&1        &2&7     &1&9        &1&7   \\
                                  &0&9       &14&0        &1&7     &0&5        &0&7   \\
                                  &1&1        &6&2        &1&1     &0&4        &0&5   \\
                                  &1&3       &3&22       &0&71    &0&44       &0&19   \\
                                  &1&5       &1&36       &0&45    &0&18       &0&22   \\\hline
\multirow{10}{*}{-1.0}            &0&1       &14&7        &1&2     &1&3        &1&1   \\
                                  &0&3       &28&2        &1&7     &1&4        &1&1   \\
                                  &0&5       &27&7        &1&7     &1&4        &1&2   \\
                                  &0&7       &20&9        &1&3     &0&9        &0&8   \\
                                  &0&9      &12&16       &0&89    &0&89       &0&46   \\
                                  &1&1       &6&96       &0&64    &0&26       &0&22   \\
                                  &1&3       &2&93       &0&39    &0&15       &0&12   \\
                                  &1&5       &1&80       &0&30    &0&12       &0&13   \\
                                  &1&7       &0&59       &0&16    &0&04       &0&04   \\
                                  &1&9       &0&38       &0&14    &0&04       &0&02   \\\hline
\multirow{10}{*}{-0.5}            &0&1      &10&74       &0&59    &1&11       &0&37   \\
                                  &0&3      &24&31       &0&95    &2&34       &0&85   \\
                                  &0&5       &25&5        &1&0     &2&2        &0&8   \\
                                  &0&7      &20&68       &1&00    &1&41       &0&72   \\
                                  &0&9      &12&05       &0&77    &0&96       &0&42   \\
                                  &1&1       &6&61       &0&55    &0&28       &0&23   \\
                                  &1&3       &3&74       &0&41    &0&18       &0&13   \\
                                  &1&5       &1&62       &0&25    &0&12       &0&09   \\
                                  &1&7       &0&87       &0&19    &0&07       &0&05   \\
                                  &1&9       &0&42       &0&14    &0&06       &0&06   \\\hline
\multirow{10}{*}{0.0}             &0&1      &10&06       &0&56    &0&47       &0&42   \\
                                  &0&3      &22&89       &0&87    &1&75       &0&61   \\
                                  &0&5      &23&26       &0&91    &1&83       &0&73   \\
                                  &0&7      &18&58       &0&89    &1&44       &0&75   \\
                                  &0&9      &11&50       &0&78    &0&66       &0&63   \\
                                  &1&1       &5&63       &0&59    &0&20       &0&33   \\
                                  &1&3       &3&74       &0&49    &0&22       &0&22   \\
                                  &1&5       &1&45       &0&27    &0&09       &0&08   \\
                                  &1&7       &0&69       &0&20    &0&05       &0&04   \\
                                  &1&9       &0&34       &0&11    &0&06       &0&08   \\\hline
\multirow{10}{*}{0.5}             &0&1      &10&79       &0&63    &1&55       &0&24   \\
                                  &0&3      &23&89       &0&95    &4&07       &0&66   \\
                                  &0&5       &27&4        &1&0     &3&9        &0&6   \\
                                  &0&7      &18&03       &0&88    &1&94       &0&51   \\
                                  &0&9      &11&04       &0&78    &1&13       &0&36   \\
                                  &1&1       &5&87       &0&57    &0&27       &0&15   \\
                                  &1&3       &3&50       &0&49    &0&14       &0&34   \\
                                  &1&5       &1&47       &0&30    &0&14       &0&15   \\
                                  &1&7       &0&65       &0&21    &0&09       &0&05   \\
                                  &1&9       &0&45       &0&14    &0&46       &0&01   \\\hline
\multirow{8}{*}{1.0}              &0&1      &11&91       &0&91    &1&55       &0&68   \\
                                  &0&3       &26&6        &1&4     &3&2        &1&2   \\
                                  &0&5       &29&0        &1&5     &6&6        &1&2   \\
                                  &0&7       &21&2        &1&4     &2&1        &1&1   \\
                                  &0&9       &13&4        &1&1     &1&7        &0&6   \\
                                  &1&1       &7&65       &0&80    &2&26       &0&11   \\
                                  &1&3       &3&32       &0&56    &0&20       &0&29   \\
                                  &1&5       &1&06       &0&28    &0&00       &0&00   \\\hline
\end{tabular}
\end{center}
\end{table}

\begin{table}
\caption{Double-differential yield $\frac{d^2n}{dydm_{_T}}$.} 

\label{tab:d2ndydmT}       
\begin{center}
\begin{tabular}{|r|r@{.}l|r@{.}l r@{.}l r@{.}l r@{.}l|}\hline
\multicolumn{1}{|r|}{$y$}&\multicolumn{2}{|l|}{$m_{_T}-$}&\multicolumn{2}{l}{$\frac{d^2n}{dydm_{_T}}$}&\multicolumn{2}{l}{$\Delta_{stat}$}
&\multicolumn{2}{l}{$\Delta_{sys}^-$}&\multicolumn{2}{l|}{$\Delta_{sys}^+$}\\
&\multicolumn{2}{l|}{$m_{_\Lambda}$}&\multicolumn{8}{l|}{$\times10^3$ ($\frac{1}{GeV/c^2}$)}\\\hline 
\multirow{7}{*}{-1.5}&      0&05       &62&0        &6&2      &8&8        &4&5  \\
                     &      0&15       &31&9        &3&2      &2&3        &1&5  \\
                     &      0&25       &18&5        &2&1      &0&8        &1&1  \\
                     &      0&35        &8&4        &1&3      &0&5        &0&3  \\
                     &      0&45       &4&51       &0&92     &0&30       &0&72  \\
                     &      0&55       &2&91       &0&73     &0&43       &0&21  \\
                     &      0&65       &1&26       &0&43     &0&15       &0&08  \\\hline
\multirow{9}{*}{-1.0}&      0&05       &53&8        &2&5      &2&4        &1&7  \\
                     &      0&15       &27&3        &1&6      &1&5        &1&1  \\
                     &      0&25       &14&6        &1&0      &0&9        &0&7  \\
                     &      0&35       &8&40       &0&72     &0&48       &0&31  \\
                     &      0&45       &5&10       &0&54     &0&25       &0&34  \\
                     &      0&55       &2&50       &0&36     &0&19       &0&08  \\
                     &      0&65       &1&44       &0&28     &0&11       &0&14  \\
                     &      0&75       &1&16       &0&24     &0&06       &0&08  \\
                     &      0&85       &0&58       &0&16     &0&05       &0&03  \\\hline
\multirow{10}{*}{-0.5}&     0&05       &45&9        &1&4      &4&3        &1&5  \\
                     &      0&15       &25&9        &1&1      &1&7        &0&8  \\
                     &      0&25      &14&56       &0&86     &1&04       &0&61  \\
                     &      0&35       &8&28       &0&63     &0&63       &0&29  \\
                     &      0&45       &4&54       &0&45     &0&21       &0&23  \\
                     &      0&55       &3&23       &0&38     &0&19       &0&10  \\
                     &      0&65       &1&83       &0&28     &0&07       &0&10  \\
                     &      0&75       &0&89       &0&18     &0&05       &0&06  \\
                     &      0&85       &0&79       &0&19     &0&09       &0&06  \\
                     &      0&95       &0&51       &0&16     &0&06       &0&07  \\\hline
\multirow{9}{*}{ 0.0}&      0&05       &43&4        &1&3      &3&0        &1&3  \\
                     &      0&15      &22&63       &0&93     &1&82       &1&19  \\
                     &      0&25      &13&38       &0&79     &0&66       &0&48  \\
                     &      0&35       &7&67       &0&67     &0&68       &0&42  \\
                     &      0&45       &4&34       &0&52     &0&20       &0&32  \\
                     &      0&55       &2&56       &0&39     &0&39       &0&14  \\
                     &      0&65       &1&89       &0&34     &0&19       &0&35  \\
                     &      0&75       &1&00       &0&24     &0&06       &0&22  \\
                     &      0&85       &0&61       &0&18     &0&09       &0&11  \\\hline
\multirow{10}{*}{ 0.5}&     0&05       &46&2        &1&4      &6&9        &7&6  \\
                     &      0&15       &25&7        &1&0      &3&5        &4&2  \\
                     &      0&25      &13&10       &0&79     &1&12       &2&15  \\
                     &      0&35       &7&28       &0&63     &0&76       &1&20  \\
                     &      0&45       &4&17       &0&48     &0&13       &0&69  \\
                     &      0&55       &3&05       &0&43     &0&17       &0&60  \\
                     &      0&65       &1&44       &0&32     &0&09       &0&27  \\
                     &      0&75       &0&77       &0&21     &0&10       &0&14  \\
                     &      0&85       &0&92       &0&25     &0&11       &0&21  \\
                     &      0&95       &0&32       &0&13     &0&03       &0&06  \\\hline
\multirow{8}{*}{ 1.0}&      0&05       &52&8        &2&1      &9&4        &8&2  \\
                     &      0&15       &26&6        &1&5      &3&0        &4&3  \\
                     &      0&25       &16&5        &1&2      &1&9        &2&7  \\
                     &      0&35       &9&04       &0&94     &1&19       &1&97  \\
                     &      0&45       &4&00       &0&61     &0&67       &0&88  \\
                     &      0&55       &2&51       &0&49     &0&31       &0&33  \\
                     &      0&65       &1&90       &0&49     &0&16       &0&46  \\
                     &      0&75       &0&63       &0&22     &0&08       &0&13  \\\hline
\end{tabular}
\end{center}
\end{table}

\newpage
\begin{table*}
\caption{Double-differential yields, $\frac{d^{2}n}{x_{_F}p_{_T}}$ and $f_n(x_{_F},p_{T})$, for $x_{_F}<0$.}
\label{tab:d2ndxFdpT}       
\begin{center}
\begin{tabular}{|r|r@{.}l|r@{.}l r@{.}l r@{.}l r@{.}l|r@{.}l r@{.}l r@{.}l r@{.}l|}\hline
\multicolumn{1}{|r|}{$x_{_F}$}&\multicolumn{2}{|l|}{$p_{_T}$}&\multicolumn{2}{c}{$\frac{d^2n}{dx_{_F}dp_{_T}}$}&\multicolumn{2}{c}{$\Delta_{stat}$}
&\multicolumn{2}{c}{$\Delta_{sys}^-$}&\multicolumn{2}{l|}{$\Delta_{sys}^+$}
&\multicolumn{2}{c}{$f_n(x_{_F},p_{_T})$}&\multicolumn{2}{c}{$\Delta_{stat}$}
&\multicolumn{2}{c}{$\Delta_{sys}^-$}&\multicolumn{2}{l|}{$\Delta_{sys}^+$}\\
&\multicolumn{2}{l|}{}&\multicolumn{8}{l|}{$\times10^3$ ($\frac{1}{GeV/c}$)}&\multicolumn{8}{l|}{$\times10^3$ $\left(\frac{1}{\left(GeV/c\right)^2}\right)$}\\\hline 
\multirow{8}{*}{-0.35}     &0&1        &\sincol{44}&       &\sincol{21}&      &\sincol{13}&        &\sincol{4}&         &\sincol{26}&        &\sincol{12}&         &\sincol{9}&         &\sincol{5}&\\ 
                          &0&3        &\sincol{127}&       &\sincol{34}&      &\sincol{30}&       &\sincol{12}&       &25&2       &6&8       &6&4       &4&7\\
                          &0&5         &\sincol{89}&       &\sincol{15}&       &\sincol{8}&       &\sincol{13}&       &10&7       &1&9       &0&6       &1&0\\
                          &0&7         &\sincol{78}&       &\sincol{12}&       &\sincol{6}&       &\sincol{10}&        &6&8       &1&0       &0&4       &0&6 \\
                           &0&9       &37&4        &6&4       &3&7        &4&0    &2&56      &0&44      &0&22      &0&19  \\
                           &1&1       &15&4        &3&4       &0&8        &1&0    &0&88      &0&19      &0&06      &0&04  \\
                           &1&3        &9&3        &2&2       &0&8        &0&1    &0&46      &0&11      &0&04      &0&02  \\
                           &1&5        &4&6        &1&5       &0&8        &0&6   &0&201     &0&066     &0&029     &0&042  \\\hline
\multirow{9}{*}{-0.25}    &0&1       &45&5         &8&9       &2&0        &3&5    &20&4       &4&0       &3&2       &3&2  \\
                           &0&3       &\sincol{110}&       &\sincol{14}&      &\sincol{11}&       &\sincol{10}&       &16&5       &2&2       &2&6       &1&1\\
                          &0&5        &\sincol{92}&        &\sincol{11}&       &\sincol{6}&        &\sincol{6}&       &8&44      &0&98      &1&14      &0&32\\
                          &0&7       &78&9         &7&4       &2&8        &2&3    &5&25      &0&49      &0&94      &0&13   \\
                           &0&9       &41&4        &4&5       &1&4        &3&9    &2&20      &0&24      &0&48      &0&06  \\
                           &1&1       &21&8        &2&8       &1&0        &1&0    &0&97      &0&13      &0&17      &0&04  \\
                           &1&3        &8&5        &1&6       &0&6        &0&6   &0&334     &0&062     &0&077     &0&012  \\
                           &1&5        &4&7        &1&1       &0&3        &0&5   &0&165     &0&038     &0&032     &0&033  \\
                           &1&7       &1&71       &0&58      &0&13       &0&17   &0&055     &0&019     &0&008     &0&005  \\\hline
\multirow{9}{*}{-0.15}     &0&1       &68&7        &5&8       &3&7        &3&0    &21&7       &1&8       &0&9       &1&5  \\
                           &0&3      &135&6        &7&8       &4&7        &5&4   &14&46      &0&83      &0&36      &0&46  \\
                           &0&5      &132&9        &7&4       &5&4        &4&9    &8&73      &0&48      &0&30      &0&22  \\
                           &0&7      &101&6        &6&0       &3&4        &4&0    &4&94      &0&29      &0&10      &0&19  \\
                           &0&9       &53&4        &3&9       &3&0        &1&9    &2&11      &0&15      &0&09      &0&05  \\
                           &1&1       &29&6        &2&7       &1&2        &1&3   &1&008     &0&092     &0&033     &0&039  \\
                           &1&3       &11&6        &1&6       &0&6        &0&5   &0&352     &0&049     &0&014     &0&011  \\
                           &1&5        &5&9        &1&1       &0&6        &0&4   &0&165     &0&030     &0&016     &0&015  \\
                           &1&7       &2&64       &0&73      &0&20       &0&11   &0&069     &0&019     &0&005     &0&002  \\\hline
\multirow{10}{*}{-0.05}    &0&1       &72&7        &3&4       &2&8        &2&5   &16&08      &0&75      &1&05      &0&26  \\
                           &0&3      &161&4        &5&4       &6&9        &4&5   &12&23      &0&41      &0&43      &0&26  \\
                           &0&5      &160&8        &5&6       &6&3        &5&4    &7&68      &0&27      &0&23      &0&24  \\
                           &0&7      &118&5        &5&2       &4&4        &3&5    &4&33      &0&19      &0&11      &0&07  \\
                           &0&9       &64&1        &4&0       &2&6        &2&2    &1&97      &0&12      &0&06      &0&05  \\
                           &1&1       &33&0        &2&9       &1&2        &1&5   &0&899     &0&078     &0&024     &0&036  \\
                           &1&3       &20&2        &2&3       &1&4        &1&0   &0&506     &0&059     &0&031     &0&036  \\
                           &1&5        &7&2        &1&2       &0&5        &0&4   &0&169     &0&029     &0&016     &0&008  \\
                           &1&7        &4&8        &1&2       &0&3        &0&3   &0&108     &0&027     &0&010     &0&006  \\
                           &1&9       &0&55       &0&32      &0&07       &0&06  &0&0120    &0&0070    &0&0016    &0&0032  \\\hline
\end{tabular}
\end{center}
\end{table*}

\begin{table*}
\caption{Double-differential yields, $\frac{d^{2}n}{x_{_F}p_{_T}}$ and $f_n(x_{_F},p_{T})$, for $x_{_F}>0$.}
\label{tab:2d2ndxFdpT}       
\begin{center}
\begin{tabular}{|r|r@{.}l|r@{.}l r@{.}l r@{.}l r@{.}l|r@{.}l r@{.}l r@{.}l r@{.}l|}\hline
\multicolumn{1}{|r|}{$x_{_F}$}&\multicolumn{2}{|l|}{$p_{_T}$}&\multicolumn{2}{c}{$\frac{d^2n}{dx_{_F}dp_{_T}}$}&\multicolumn{2}{c}{$\Delta_{stat}$}
&\multicolumn{2}{c}{$\Delta_{sys}^-$}&\multicolumn{2}{l|}{$\Delta_{sys}^+$}
&\multicolumn{2}{c}{$f_n(x_{_F},p_{_T})$}&\multicolumn{2}{c}{$\Delta_{stat}$}
&\multicolumn{2}{c}{$\Delta_{sys}^-$}&\multicolumn{2}{l|}{$\Delta_{sys}^+$}\\
&\multicolumn{2}{l|}{}&\multicolumn{8}{l|}{$\times10^3$ ($\frac{1}{GeV/c}$)}&\multicolumn{8}{l|}{$\times10^3$ $\left(\frac{1}{\left(GeV/c\right)^2}\right)$}\\\hline 
\multirow{10}{*}{ 0.05}  &0&1       &70&3        &3&5         &8&9        &2&7        &15&56      &0&77      &2&49      &1&11  \\
                         &0&3      &160&9        &5&4        &22&8        &3&7        &12&20      &0&41      &1&59      &0&15  \\
                         &0&5      &166&4        &5&6        &20&8        &4&0         &7&95      &0&27      &0&97      &0&09  \\
                         &0&7      &113&1        &4&9        &13&3        &2&9         &4&13      &0&18      &0&45      &0&08  \\
                         &0&9       &64&4        &4&2         &7&4        &1&7         &1&97      &0&13      &0&16      &0&04  \\
                         &1&1       &33&5        &3&2         &2&5        &1&0        &0&913     &0&086     &0&065     &0&028  \\
                         &1&3       &17&0        &2&3         &1&1        &0&8        &0&426     &0&058     &0&035     &0&017  \\
                         &1&5        &6&0        &1&3         &0&2        &0&4        &0&141     &0&030     &0&004     &0&009  \\
                         &1&7       &2&56       &0&82        &0&33       &0&14        &0&058     &0&018     &0&007     &0&003  \\
                         &1&9       &1&19       &0&51        &0&21       &0&23        &0&026     &0&011     &0&005     &0&006  \\\hline
\multirow{9}{*}{ 0.15}   &0&1       &63&1        &4&5         &8&5        &3&4         &19&9       &1&4       &4&3       &1&0  \\
                         &0&3      &131&5        &6&6        &26&7        &5&1        &14&02      &0&70      &2&28      &0&71  \\
                         &0&5      &146&8        &7&0        &35&4        &3&5         &9&64      &0&46      &1&79      &0&35  \\
                         &0&7       &87&8        &5&7        &16&2        &3&4         &4&27      &0&28      &0&54      &0&19  \\
                         &0&9       &61&2        &4&7         &6&7        &1&8         &2&42      &0&19      &0&62      &0&09  \\
                         &1&1       &22&3        &2&7         &2&4        &1&1        &0&758     &0&091     &0&077     &0&039  \\
                         &1&3       &15&1        &2&5         &2&2        &1&3        &0&460     &0&077     &0&039     &0&043  \\
                         &1&5        &6&6        &1&4         &1&0        &0&4        &0&185     &0&039     &0&032     &0&012  \\
                         &1&7       &1&39       &0&63        &0&14       &0&29        &0&036     &0&017     &0&007     &0&007  \\\hline
\multirow{8}{*}{ 0.25}   &0&1       &43&7        &7&0         &9&4        &5&2         &19&6       &3&1       &2&6       &2&6  \\
                         &0&3       &\sincol{115}&      &\sincol{10}&          &\sincol{8}&        &\sincol{8}&            &17&3       &1&5       &3&4       &1&2  \\
                         &0&5      &104&7        &9&4        &11&1        &4&7         &9&57      &0&86      &1&00      &1&20  \\
                         &0&7       &91&9        &8&2        &13&4        &5&4         &6&12      &0&55      &1&58      &0&41  \\
                         &0&9       &44&7        &5&8         &8&9        &2&7         &2&37      &0&31      &0&46      &0&09  \\
                         &1&1       &16&8        &3&3         &3&5        &1&5         &0&75      &0&15      &0&17      &0&05  \\
                         &1&3        &8&7        &2&1         &0&7        &0&8        &0&341     &0&082     &0&039     &0&045  \\
                         &1&5        &2&9        &1&1         &0&4        &0&4        &0&102     &0&037     &0&015     &0&010  \\\hline
\multirow{7}{*}{ 0.35}   &0&1        &\sincol{29}&      &\sincol{12}&         &\sincol{16}&       &\sincol{14}&            &17&5       &7&4       &4&3       &1&7  \\
                         &0&3        &\sincol{81}&      &\sincol{17}&         &\sincol{28}&       &\sincol{23}&            &16&1       &3&4       &6&4       &4&6  \\
                         &0&5        &\sincol{91}&      &\sincol{16}&         &\sincol{39}&       &\sincol{13}&            &10&9       &1&9       &1&8       &2&4  \\
                         &0&7        &\sincol{61}&      &\sincol{11}&         &\sincol{25}&       &\sincol{19}&            &5&28      &0&95      &0&81      &1&79  \\
                         &0&9       &37&4        &8&2        &22&3        &6&0         &2&56      &0&56      &0&48      &0&26  \\
                         &1&1       &17&3        &5&1         &6&3        &6&0         &0&99      &0&29      &0&29      &0&13  \\
                         &1&3       &11&7        &3&6         &4&3        &2&6         &0&57      &0&18      &0&18      &0&14  \\\hline
\end{tabular}
\end{center}
\end{table*}

\begin{table*}
\caption{$p_{_T}$ integrated yield $\frac{dn}{dy}$, the inverse slope parameter $T$ and the  mean transverse mass $\langle m_{_T}\rangle - m_{\Lambda}$.}

\label{tab:dndy}       
\begin{center}
\begin{tabular}{|r@{.}l|r@{.}l r@{.}l r@{.}l r@{.}l|r@{.}l r@{.}l r@{.}l r@{.}l|r@{.}l r@{.}l r@{.}l r@{.}l|}\hline
\multicolumn{2}{|l|}{y}&\multicolumn{2}{c}{$\frac{dn}{dy}$}&\multicolumn{2}{l}{$\Delta_{stat}$}
&\multicolumn{2}{l}{$\Delta_{sys}^-$}&\multicolumn{2}{l|}{$\Delta_{sys}^+$}
&\multicolumn{2}{l}{$T$}&\multicolumn{2}{l}{$\Delta_{stat}$}&\multicolumn{2}{l}{$\Delta_{sys}^-$}&\multicolumn{2}{l|}{$\Delta_{sys}^+$}
&\multicolumn{2}{l}{$\langle m_{_T}\rangle - m_{\Lambda}$}&\multicolumn{2}{l}{$\Delta_{stat}$}&\multicolumn{2}{l}{$\Delta_{sys}^-$}&\multicolumn{2}{l|}{$\Delta_{sys}^+$}\\
\multicolumn{2}{|l|}{}&\multicolumn{2}{l}{$\times10^3$}&\multicolumn{6}{c|}{} &\multicolumn{2}{c}{(MeV)}&\multicolumn{6}{c|}{}&\multicolumn{2}{c}{($\frac{GeV}{c^2}$)}&\multicolumn{6}{c|}{}\\\hline 
      -1&5  &26&8      &1&5      &2&4        &1&4        &143&8      &6&3    &5&4        &3&3    &0&156    &0&013      &0&005      &0&006\\
      -1&0  &23&30     &0&65    &1&02       &0&73        &152&8      &3&8    &4&2        &4&1   &0&1687   &0&0076     &0&0050     &0&0046\\
      -0&5  &21&35     &0&43    &1&71       &0&64        &163&0      &3&2    &4&5        &5&1   &0&1813   &0&0067     &0&0050     &0&0056\\
       0&0  &19&65     &0&40    &1&14       &0&60        &160&7      &3&6    &4&3        &5&2   &0&1777   &0&0076     &0&0051     &0&0068\\
       0&5  &20&64     &0&42    &2&53       &0&43        &154&0      &3&6    &3&9        &8&4   &0&1697   &0&0070     &0&0037     &0&0102\\
       1&0  &22&98     &0&62    &2&96       &0&65        &153&9      &4&1    &4&0        &4&6   &0&1640   &0&0085     &0&0028     &0&0084\\\hline
\end{tabular}
\end{center}
\end{table*}

\begin{table*}
\caption{$p_{_T}$ integrated yield $\frac{dn}{x_{_F}}$ and the invariant cross section $F(x_{_F})$.}
\label{tab:dndxF}       
\begin{center}
\begin{tabular}{|r@{.}l|r@{.}l r@{.}l r@{.}l r@{.}l|r@{.}l r@{.}l r@{.}l r@{.}l|}\hline
\multicolumn{2}{|l|}{$x_{_F}$}&\multicolumn{2}{c}{$\frac{dn}{dx_{_F}}$}&\multicolumn{2}{l}{$\Delta_{stat}$}
&\multicolumn{2}{l}{$\Delta_{sys}^-$}&\multicolumn{2}{l|}{$\Delta_{sys}^+$} 
&\multicolumn{2}{c}{$F(x_{_F})$}&\multicolumn{2}{l}{$\Delta_{stat}$}
&\multicolumn{2}{l}{$\Delta_{sys}^-$}&\multicolumn{2}{l|}{$\Delta_{sys}^+$}\\ 
\multicolumn{2}{|l|}{}&\multicolumn{2}{l}{$\times10^3$}&\multicolumn{6}{l|}{}&\multicolumn{3}{l}{$\times10^3$~(mb)}&\multicolumn{5}{l|}{}\\\hline
     -0&35     &81&3      &9&0            &9&3        &6&5      &\sincol{313}&       &\sincol{34}&       &\sincol{36}&       &\sincol{70}&\\   
     -0&25     &81&1      &4&4            &3&9        &4&7      &\sincol{239}&       &\sincol{13}&       &\sincol{10}&       &\sincol{13}&\\   
     -0&15     &108&6      &2&9           &3&9        &3&5    &231&6       &6&1       &5&9       &4&9 \\                                   
     -0&05     &128&7      &2&3           &4&7        &3&7    &204&1       &3&7       &5&2       &4&5 \\                                   
      0&05      &127&2      &2&3         &14&4        &3&0    &200&9       &3&7      &22&3       &2&2 \\                                   
      0&15      &107&3      &2&7         &18&5        &3&0    &228&9       &5&8      &39&2       &6&7 \\                                   
      0&25      &86&0      &3&8           &9&8        &4&0      &\sincol{253}&       &\sincol{11}&       &\sincol{29}&       &\sincol{12}&\\   
      0&35      &67&0      &6&1          &25&8       &11&6      &\sincol{258}&       &\sincol{23}&       &\sincol{99}&       &\sincol{45}&\\   \hline
\end{tabular}
\end{center}
\end{table*}

\end{document}